%% file: KeldyshRevised.tex
\documentclass{article}
\newfont{\bb}{msbm10}

\def\Frac{\displaystyle \frac}
\usepackage{amsfonts}
\usepackage{amsmath,amssymb}
\usepackage{graphicx}
\usepackage{color}
\usepackage{caption}                            %
\usepackage{subcaption}                 

\def\Div{{\rm div}\,}
\newcommand{\bu} {\mathbf{u}}
\newcommand{\R} {\mathbb{R}}

\newcommand{\bn} {\mathbf{n}}

\newcommand{\bx} {\mathbf{x}}
\newcommand{\ba} {\mathbf{a}}
\newcommand{\bX} {\mathbf{X}}

\newcommand{\by} {\mathbf{y}}
\newcommand{\bp} {\mathbf{p}}
\newcommand{\bt} {\mathbf{t}}

\newcommand{\bg} {\mathbf{g}}

\newcommand{\T} {{\cal T}}
\newcommand{\F} {{\cal F}}
\newcommand{\N} {{\cal N}}
\newcommand{\cP} {{\cal P}}

\newtheorem{remark}{Remark}

\usepackage{cite}

\renewcommand{\phi}{\varphi}

\begin{document}


\title{An adaptive numerical method for free surface flows passing rigidly mounted obstacles\thanks{Supported by Russian Science Foundation through the grant 14-11-00434}} 

\author{
        Kirill~D.~Nikitin~\thanks{Institute of Numerical Mathematics of the Russian Academy of Sciences, Moscow, Russia and Keldysh Institute of Applied Mathematics of the Russian Academy of Sciences, Moscow, Russia} \and Maxim~A.~Olshanskii\thanks{Department of Mathematics, University of Houston, Houston, TX, USA} \and Kirill~M.~Terekhov~\thanks{School of Earth, Energy \& Environmental Sciences, Stanford University, Stanford, CA, USA} \and
        Yuri~V.~Vassilevski~\thanks{Institute of Numerical Mathematics of the Russian Academy of Sciences, Moscow, Russia and Keldysh Institute of Applied Mathematics of the Russian Academy of Sciences, Moscow, Russia} \and Ruslan~M.~Yanbarisov~\thanks{Moscow Institute of Physics and Technology, Dolgoprudny, Russia}
}



\date{}
\maketitle

\begin{abstract}
The paper develops a method for the numerical simulation of a free-surface flow of incompressible viscous fluid around a streamlined body. The body is a rigid stationary construction partially submerged in the fluid. The application we are interested in the paper is a flow around a surface mounted offshore oil platform. The numerical method builds on a hybrid finite volume\,/\,finite difference discretization using adaptive octree cubic meshes. The mesh is dynamically refined towards the free surface and the construction. Special care is taken to devise a discretization for the case of curvilinear boundaries and interfaces immersed in the octree Cartesian background computational mesh. To demonstrate the accuracy of the method, we show the results for two benchmark problems: the sloshing 3D container and the channel laminar flow passing the 3D cylinder of circular cross-section. Further, we simulate numerically a flow with surface waves around an offshore oil platform for the realistic set of geophysical  data.
\end{abstract}

{\bf Keywords:}
free surface, incompressible flow, mesh adaptation, Navier-Stokes, octree meshes, curvilinear boundaries, sloshing container, 3D cylinder of circular cross-section, flow around oil platform



\section{Introduction}\label{intro}
Free surface flows passing partially submerged objects are common in nature and engineering applications. The examples include water flows around bridge piers, ship bodies, water plants, or costal constructions.
A mathematical model of such phenomena includes fluid dynamics equations and an evolution equation for the free surface. These equations can be posed in a domain of complex geometry. Handling the equations and the geometry  numerically in an efficient and accurate way constitutes the major challenge for a CFD method applied to simulate free surface flows passing submerged obstacles. Depending on the applications, the fluid and free surface equations can be coupled to other mathematical models of transport, elasticity, etc. Thus, a reliable fast and accurate solver is desirable.

The previous studies of free surface flows passing submerged bodies include the simulation of Euler flows around hydrofoils \cite{Hino_1988}, a boundary element method with the Lagrangian treatment of free surface evolution \cite{Kang_2015},
a non-body conformal grid finite difference method for compressible flows \cite{Ghias_2004},
a stabilized finite element method for fluid equations in ALE form \cite{Onate2001635},
and other FEM-based ALE techniques for fluid-structure interaction described in \cite{Bazilevs_2013}.
The variants of the immersed boundary method \cite{Peskin_1972,Mittal_2005} for the free surface flows were discussed in \cite{Zhang_2010,Liu_2014}.
Analytical and semi-analytical solutions of the free surface flows around specific submerged bodies were studied in \cite{Tyvand_1995,Chatjigeorgiou_2014}.

The method developed in this paper is based on a hybrid  discretization using octree Cartesian background meshes.
Octree meshes enjoy a growing reliance in scientific computing community due to the simple Cartesian structure and embedded
hierarchy, which makes mesh adaptation, reconstruction and data access fast and easy. In particular, octree meshes can be dynamically adapted towards the free surface. The adaptation can be also based on various error indicators.   Fast remeshing with octree grids makes them a natural choice for the simulation of moving interfaces and free surface flows, see, e.g., \cite{Sochnikov,Losasso:04,Losasso:07,Pop09,Faster_etal,NikitinVassilevski:08}, as well as more general non-Newtonian and high-speed Newtonian flows, see, e.g., \cite{bonito2016numerical,BR2006,afkhamiPoF15a,Olshanskii2013231,Popinet2003572,Zingan2013479}.
The Cartesian structure of octree meshes requires, however, a special technique for handling curvilinear boundaries and interfaces, since the mesh itself provides only the first order geometric accuracy in this case.

Using octree grids for the simulation of flows over partially submerged bodies  gives the advantage of better local resolution of the free surface and fluid interaction with the body. For the more accurate treatment of the equations near the curvilinear boundary of the construction, we immerse the rigid object in the background mesh and construct the second order approximation of the fluid and free surface equations in the cut cells. The level-set method is used to recover the evolution of the free surface. Other  important ingredients of our approach are the semi-Lagrangian characteristic method for the level-set equations  on the dynamic octree meshes from \cite{terekhov2015semi}, and the splitting  method for the fluid equations on the  octree meshes from \cite{Olshanskii2013231} with  filtering. In that paper, the method was studied for enclosed incompressible viscous flows in cavities and over bluff bodies.

Compared to well-studied higher-order {\color{black}finite volume and finite difference} discretizations on uniform grids,  the schemes that exploit adaptivity properties of  octree meshes  often pay the price of lower accuracy and higher numerical dissipation. This happens due to the presence of
hanging nodes {\color{black} on irregular interfaces and non-uniform mesh size, which require interpolation of unknowns and make impossible certain cancellations of discretization errors. Such error cancellations take place for uniform grid due to the stencil symmetry}. 
To overcome this loss of accuracy, we operate with a suitable sets of nodes and least-square minimizing interpolants. Further, we validate our approach by performing a series of numerical experiments.
First, we compute a channel flow past a 3D circular cylinder.  
Second, we simulate the sloshing of water in a 3D tank subject to periodic horizontal excitation.
The critical statistics, which are drag, lift coefficients for the first test and water levels for the second test,
are compared against reference data found in the literature.
The success of the numerical method for both benchmark problems demonstrates its ability to accurately simulate incompressible viscous free-surface flows and flows passing streamlined bodies with curvilinear boundaries. Therefore, we apply  the method to simulate
the water flow with surface waves around an offshore oil platform rigidly mounted in the Kara sea offshore.
The platform is a reconstruction of a currently operating unit.  
The sea waves runup reproduces the realistic weather scenario in the region of the Kara sea offshore.
The statistics of interest are water levels at the platform and forces experienced by the construction.

The rest of the paper is organized as follows. Section~\ref{sec:model} reviews the mathematical model.
Section~\ref{sec:split} presents the splitting method for the numerical time integration. Section~\ref{sec:spartial} discusses the details of the  discretization on the gradely refined  octree meshes. In section~\ref{sec:curv} we devise the numerical treatment of the curvilinear boundaries embedded in the background mesh. Section~\ref{sec:num} collects the results of numerical experiments.

\section{Mathematical model}\label{sec:model}

Consider a Newtonian incompressible fluid flow  in a bounded  time-dependent domain
 $\Omega(t)\in\mathbb{R}^3$ for  $t\in(0,T]$.
The fluid dynamics is governed by the incompressible Navier-Stokes equations
\begin{equation}\label{eq:ns}
\left\{
\begin{split}
\rho\left(\frac{\partial {\mathbf{u}}}{\partial t}+ ({\mathbf{u}} \cdot \nabla) {\mathbf{u}}\right) - \Div \boldsymbol{\sigma}(\bu,p) = \bg\\
\nabla \cdot {\mathbf{u}} = 0
\end{split}\right. \quad\text{in}~ \Omega(t),~t\in(0,T],
\end{equation}
where $\boldsymbol{\sigma}(\bu,p)=\nu [\nabla \bu + (\nabla \bu)^T]- p\,\mathbf{I}$ is the stress tensor of the fluid,
${\mathbf{u}}$ is the  velocity vector field, $p$ is the kinematic  pressure, $\bg$ is the external force (e.g., gravity), $\rho$ is the density, and $\nu$ is the kinematic viscosity. At the initial time $t=0$ the domain and the velocity
field are known:
\begin{equation}\label{eq:initial}
\Omega(0)=\Omega_0,\quad \mathbf{u}|_{t=0} = \mathbf{u}_0,~~\nabla\cdot \mathbf{u}_0=0.
\end{equation}
 We assume that $\overline{\partial \Omega(t)}=\overline{\Gamma_D}\cup\overline{\Gamma(t)}\cup\overline{\Gamma}_{\rm out}\cup\overline{\Gamma}_{\rm in}$,
where $\Gamma_D$ is  the static boundary(walls), $\Gamma(t)$  is the free surface of fluid, $\Gamma_{\rm in}$, $\Gamma_{\rm out}$ are inflow and outflow parts of the boundary, respectively. Note, that $\Gamma_D$, $\Gamma_{\rm in}$, $\Gamma_{\rm out}$  may vary in time, in general.
 We assume the free surface $\Gamma(t)$  passively evolves with the normal velocity of fluid, i.e., the
following kinematic condition is valid
\begin{equation}\label{eq:nvel}
v_{\Gamma} = \mathbf{u} \cdot {\color{black}\bn}\quad \text{on}~\Gamma(t),
\end{equation}
where $\mathbf{n}$ is the normal vector for  $\Gamma(t)$ and $v_{\Gamma}$ is the normal velocity of $\Gamma(t)$.
{\color{black}Since the free surface flows we interested in this paper have large Weber numbers,  we ignore the capillary forces and}
the boundary condition on $\Gamma(t)$ reads
\begin{equation}\label{eq:tension}
\boldsymbol{\sigma}(\bu,p)  \mathbf{n} = \mathbf{0} \quad\text{on}~ \Gamma(t).
\end{equation}

On the static part of the flow boundary,  we assume the velocity field satisfies either no-slip boundary condition
\begin{equation}\label{eq:bc}
\mathbf{u}=\mathbf{0}\quad\text{on}~ \Gamma_D,
\end{equation}
or no-penetration and free-slip boundary conditions:
\begin{equation}\label{eq:bc2}
\mathbf{u}\cdot\bn=\mathbf{0}\quad \text{and}\quad \frac{\partial(\mathbf{u}\cdot\bt_i)}{\partial \bn} =0,~i=1,2, \quad\text{on}~ \Gamma_D,
\end{equation}
where $\bt_i$ and $\bn$ are tangential and normal vectors on $\Gamma_D$.
We shall use the generic notation $\mathcal{B}\bu|_{\Gamma_D}$ to denote boundary conditions \eqref{eq:bc} or \eqref{eq:bc2} on $\Gamma_D$.
We assume that $\mathbf{u}$ is given on $\Gamma_{\rm in}$ and $\boldsymbol{\sigma}(\bu,p)\bn=\mathbf{0}$ on $\Gamma_{\rm out}$.

For computational purposes, we shall employ  the implicit definition of the free surface
evolution with the help of an indicator function. Let $\Gamma(t)$ be given as the zero  level of a globally defined Lipschitz continuous \textit{level set}  function $\phi(t,\bx)$  such that
\[
\phi(t,\bx) =
\begin{cases}
    <0 & \mbox{if}~ \bx\in\Omega(t) \cr
    >0 & \mbox{if}~ \bx\in \mathbb{R}^3\setminus\overline{\Omega(t)}  \cr
    =0 & \mbox{if}~ \bx\in \Gamma(t)
\end{cases}\qquad \mbox{for all}~t\in [0,T].
\]
The initial condition \eqref{eq:initial} defines $\phi(0,\bx)$.
The kinematic condition \eqref{eq:nvel} implies that for $t>0$ the level set function can be  found as the solution to the transport equation \cite{Fedkiw:02}:
\begin{equation}
\label{eq:ls}
\frac{\partial\phi}{\partial t} + {\widetilde{\mathbf{u}}} \cdot \nabla \phi = 0 \quad \text{in}~\mathbb{R}^3\times(0,T],
\end{equation}
where $\widetilde{\mathbf{u}}$ is any (divergence-free) smooth velocity field such that $\widetilde{\mathbf{u}}=\mathbf{u}$ on $\Gamma(t)$.

 A numerical method studied in this paper solves the system of equations, boundary and initial conditions \eqref{eq:ns}--\eqref{eq:ls}.
The implicit definition of $\Gamma(t)$ as  zero level of a globally defined function $\phi$
leads to numerical algorithms which can easily handle complex topological changes of the free surface. The level set function provides an easy access to useful  geometric characteristics
of $\Gamma(t)$. For instance,  the unit outward normal to $\Gamma(t)$ is
${\color{black}\mathbf{n}} = \nabla \phi/ | \nabla \phi |$, and the
surface curvature is $\kappa = \nabla \cdot {\color{black}\mathbf{n}}$.
From the numerical point of view, it is often beneficial  if the level set function possesses the signed distance property, i.e. it satisfies  the Eikonal equation
\begin{equation}\label{eq:Eik}
| \nabla \phi|=1.
\end{equation}

\section{Numerical method} The section describes the key  ingredients of our numerical approach.

\subsection{Numerical time integration}\label{sec:split}
We consider  a semi-implicit spitting method based on the semi-Lagrangian approach for the level-set function evolution and
a hybrid finite volume\,/\,finite difference solvers for the convection-diffusion equations and the Poisson equation for pressure. The algorithm
is built on the well-known splitting procedure due to Chorin, Yanenko,  Pironneau and others, see, for example,  \cite{Chorin:68,Pironneau}.
For the sake of presentation simplicity, in this section we ignore the spacial discretization.
Important implementation details and the spacial discretization will be addressed in the next section. \smallskip

We adopt the notation $\bu^n$, $p^n$, $\phi^n$ for approximations to the velocity field, the pressure, and the level set function  at $t=t_n$.
Function $\phi^n$ implicitly defines an approximation to fluid domain at time $t=t_n$ through
$
\Omega_n:=\{\bx\in\R^3\,:\,\phi^n(\bx)<0\}.
$

Initial conditions define $\bu^0=\bu(t_0)$ and  $\phi^0=\phi(t_0)$.
For $n=0,1,\dots$ and given  $\bu^n$,  $\phi^n$ such that $\Div\bu^n=0$, we find
$\bu^{n+1}$, $p^{n+1}$, $\phi^{n+1}$ in several steps:
\smallskip

\noindent\textit{The semi-Lagrangian step: $\Omega_n\rightarrow \Omega_{n+1}$.}
Consider {\color{black} the closest-point extension of the velocity at the boundary} to the exterior of fluid domain: $\bu^n|_{\Omega_n}\to{\bu}^n|_{\mathbb{R}^3}$.
     In practice, the extension is performed to a bulk computational domain, rather than $\mathbb{R}^3$.
    For every  $\by\in\mathbb{R}^3$, solve the characteristic equation backward in time
\begin{equation}\label{eq:Char}
\frac{\partial \bx(\tau)}{\partial \tau}= \widetilde{\bu}^n(\bx(\tau)),\quad \bx(t_{n+1})=\by,\quad\text{for}~\tau\in[t_{n+1},t_n].
\end{equation}
The mapping $\bX:\by\to\bx(t_n)$ defines an isomorphism on $\R^3$.
Now,  set
\begin{equation}\label{eq:sL}
\phi^{n+1}(\by)=\phi^n(\bX(\by)).
\end{equation}
For the numerical integration of \eqref{eq:Char} we apply the  trapezoidal rule
\begin{equation}
\mathbf{x}(t_n+\frac{\Delta t}{2}) = \mathbf{x}_0 - \frac{\Delta t}{2} \mathbf{u}(\mathbf{x}_0,t_n), \quad \mathbf{x}(t_n) = \mathbf{x}_0 - \Delta t \widetilde{\mathbf{u}}^{n+\frac{1}{2}},
\label{eq:discr_slfwd}
\end{equation}
with $\Delta t = t_n-t_{n+1}$.  Since the velocity field is not given \textit{a priori}, but recovered numerically at times $t_k$, $k=0,\dots,n$,  the linear extrapolation is used:
\[
\widetilde{\mathbf{u}}^{n+\frac{1}{2}} = (1 + \eta) \mathbf{u}(\mathbf{x}(t_n+\Delta t/2),t_n) - \eta \mathbf{u}(\mathbf{x}(t_n+\Delta t/2),t_{n-1}),\quad \eta = \frac{t_{n+1} - t_n}{t_n - t_{n-1}}.
\]
 To improve the accuracy of the semi-Lagrangian step, we apply the back-and-forth error compensation and correction (BFECC) technique from \cite{dupont2003back,dupont2007back}: The same method is applied to integrate numerically the level-set equation forward in time to obtain an approximation to the error at time $t_n$. Further, the backward integration is performed one more time, but with the corrected level-set function values at time $t_n$. {\color{black}A tricubic interpolation is used to prescribe a value to $\phi^n$ at $\bX(\by)$.
The interpolation is not monotone; therefore, a limiter is introduced to reduce oscillations.}
{\color{black}For smooth solutions, the method demonstrated second order of convergence for dynamically reconstructed meshes.}
 Further  details of the semi-Lagrangian BFECC method with a limiter on the octree grids can be found in \cite{terekhov2015semi}.

 {\color{black} After the completion of the semi-Lagrangian step, we perform the re-initialization of the level set function to satisfy equation~\eqref{eq:Eik}. For this purpose, we use an algorithm from~\cite{Nikitin2014} based on  the marching cubes method for free surface triangulation and a higher order closest point method. The numerical integration of \eqref{eq:Char} may also cause a divergence (loss or gain) of the fluid volume. So we perform the volume correction with the help of the procedure described in~\cite{Nikitin2014}. We  note that the use of the BFECC method makes the re-initialization and volume correction steps less critical compared to the standard linear semi-Lagrangian method, but still they are necessary for long-time simulations.}
 \smallskip

\noindent \textit{Remeshing}. Given the new fluid domain, we update and adapt the grid to account for the new position of the free surface. The adaptation is based on the information about the  distance to the free surface provided by $\phi^{n+1}$. 
\smallskip

\noindent \textit{Re-interpolation}. After remeshing we re-interpolate all discrete variables to the new grid. The re-interpolated velocity field is defined on the bulk computational domain (due to the extension procedure at the beginning of the level-set part).
\medskip

Next we handle viscous and inertia  terms and  project the velocity into (discretely) divergence-free functions subspace and recover the new pressure. We denote $\Gamma_1=\Gamma_D\cup\Gamma_{\rm in}$, $\Gamma_2=\Gamma(t_{n+1})\cup\Gamma_{\rm out}$. \\[0.5ex]
\noindent \textit{The convection-diffusion step}: Solve for $\widetilde{\bu^{n+1}}$ in $\Omega_{n+1}$:
\begin{equation}  \label{step1}
\left\{
\begin{split}
\frac{\alpha \widetilde{\bu^{n+1}} + \beta\bu^{n}+\gamma\bu^{n-1}}{\triangle t_n} +
 (\bu^{n}+\xi(\bu^{n}-\bu^{n-1}))\cdot \nabla \widetilde {\bu^{n+1}} -
\nu\Delta \widetilde {\bu^{n+1}} & = - \nabla p^n, \\
\widetilde {\bu^{n+1}}|_{\Gamma_{\rm in}}= \bu_{\rm in},\quad \mathcal{B}\widetilde {\bu^{n+1}}|_{\Gamma_D}= \mathbf{0},\quad \left.(\nabla \widetilde{\bu^{n+1}}+\nabla \widetilde{\bu^{n+1}}^T)\bn\right|_{\Gamma_2}&=0.
\end{split}
\right.
\end{equation}
 Here $\xi = \triangle t_n / \triangle t_{n-1} $, $\alpha = 1 + \xi / (\xi+1)$, $\beta = -(\xi + 1)$, $\gamma = \xi^2 / (\xi+1)$.\\[0.5ex]
\noindent \textit{The projection step}:
Project $\widetilde{\bu^{n+1}}$ on the divergence-free space to recover $\bu^{n+1}$:
\begin{equation}  \label{step2a}
\left\{
\begin{split}
\alpha ( \bu^{n+1}- \widetilde {\bu^{n+1}}) / \triangle t_n - \nabla q & = 0, \\
\operatorname{ div} \bu^{n+1} & = {0}, \\
\bn\cdot \bu^{n+1}|_{\Gamma_1} = 0,\quad q|_{\Gamma_2}&= 0.
\end{split}
\right.
\end{equation}
The problem \eqref{step2a} is reduced to the Poisson problem for $q$:
\begin{equation}  \label{step2}
\left\{
\begin{split}
-\Delta q &= \alpha / \triangle t_n \operatorname{ div} \widetilde {\bu^{n+1}}, \\
q|_{\Gamma_2}&= 0,\quad \left.\frac{\partial q}{\partial\bn}\right|_{\Gamma_1}  = 0.
\end{split}
\right.
\end{equation}
 Finally, update the pressure:
\begin{equation}  \label{step3}
p^{n+1}=p^{n}-q+\nu\Div\widetilde{\bu^{n+1}}.
\end{equation}
The `extra' divergence term in the pressure correction step \eqref{step3} is used to reduce numerical boundary layers in the pressure, see, e.g.,~\cite{GMS,Prohl}. {\color{black}In this paper we do not address the problem of building a higher  order accurate (with respect to the time step) stable pressure projection method for the case of open boundary conditions, cf.  \cite{GMS,lee2016stability,Olshanskii2013231}.}

\subsection{Spatial discretization}\label{sec:spartial}
For the spatial discretization we use octree cubic meshes, which allow  fast dynamic mesh adaptation based on geometric or error indicators.

\begin{figure}[h!]
\begin{center}
     \includegraphics[width=0.6\linewidth]{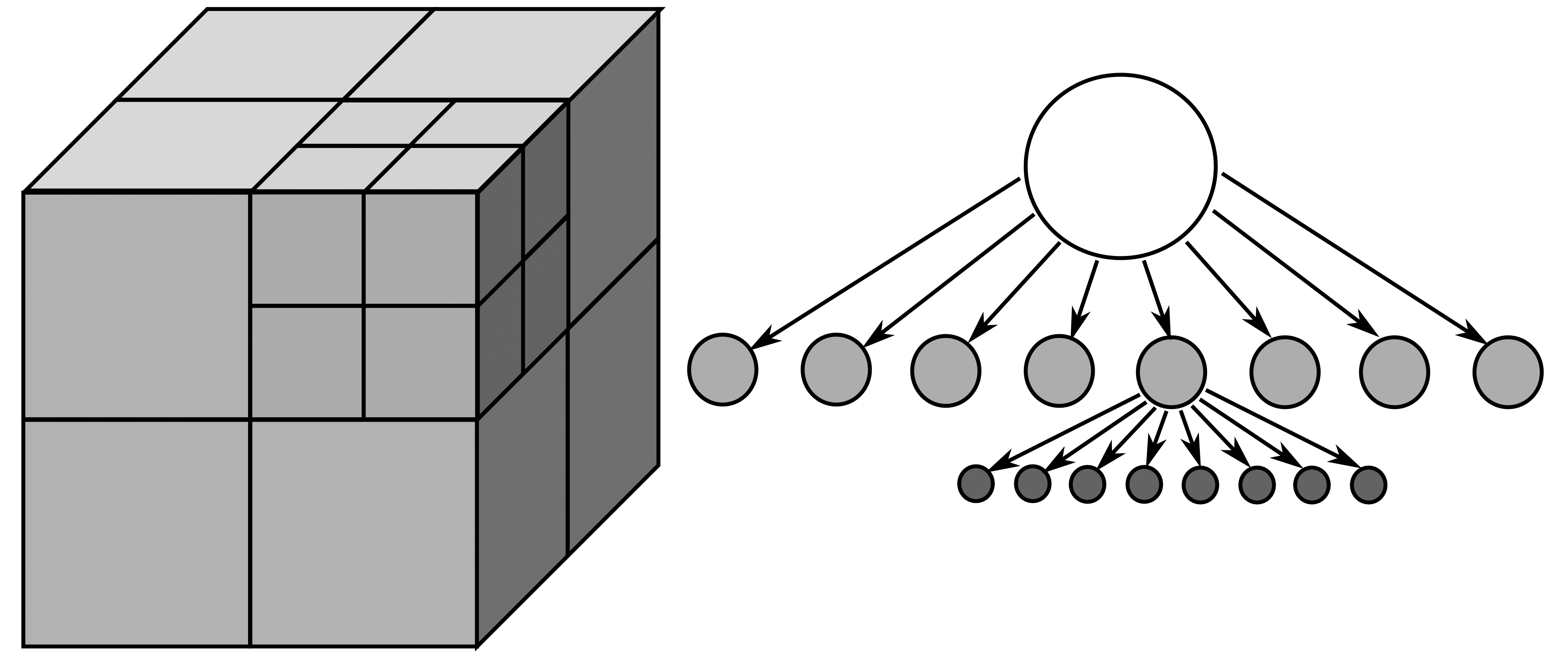}
\end{center}
\caption{An octree mesh (left) and its representation as a tree (right).}
\label{pic:octree_schematic}
\end{figure}

 Consider a graded octree  mesh with cubic cells, see   Fig.~\ref{pic:octree_schematic}.  An octree mesh is \textit{graded} if the size of  cells sharing (a part of) an edge or a face can differ in size only by the factor of two. This restriction simplifies support of mesh connectivity and the construction of discrete differential operators.
We use the staggered location of velocity and pressure unknowns. The pressure degrees of freedom are assigned to cells centers  and velocity variables are located at cells faces in such a way that every face stores normal velocity flux.
If a face is shared by  cells from different grid levels, then  velocity degrees of freedom are assigned to
the faces centers of fine grid cells (in the case of graded octree mesh, the corresponding face of the coarse grid cell holds 4 unknowns).

\begin{figure}
 \centering

 \qquad
\qquad
 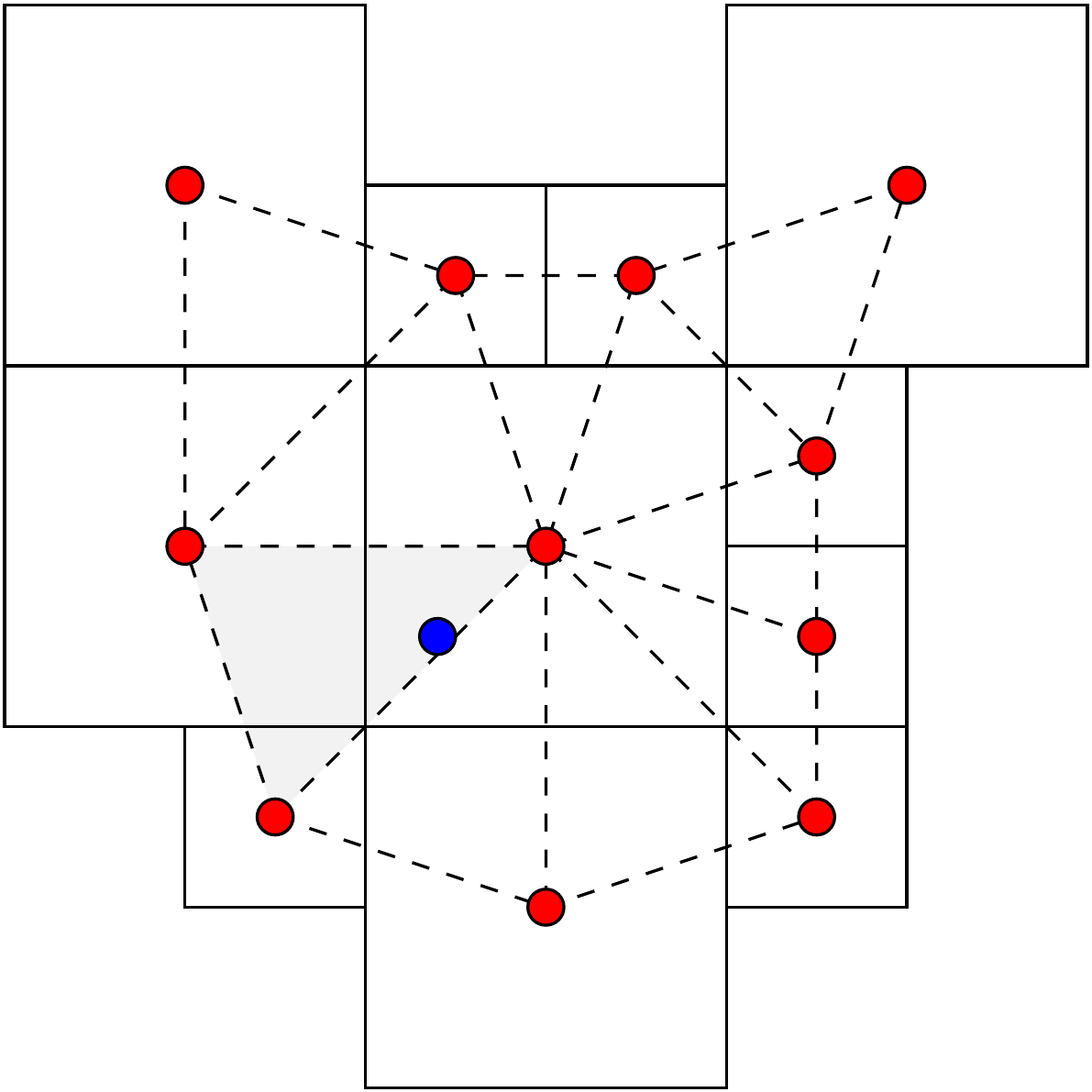\qquad
 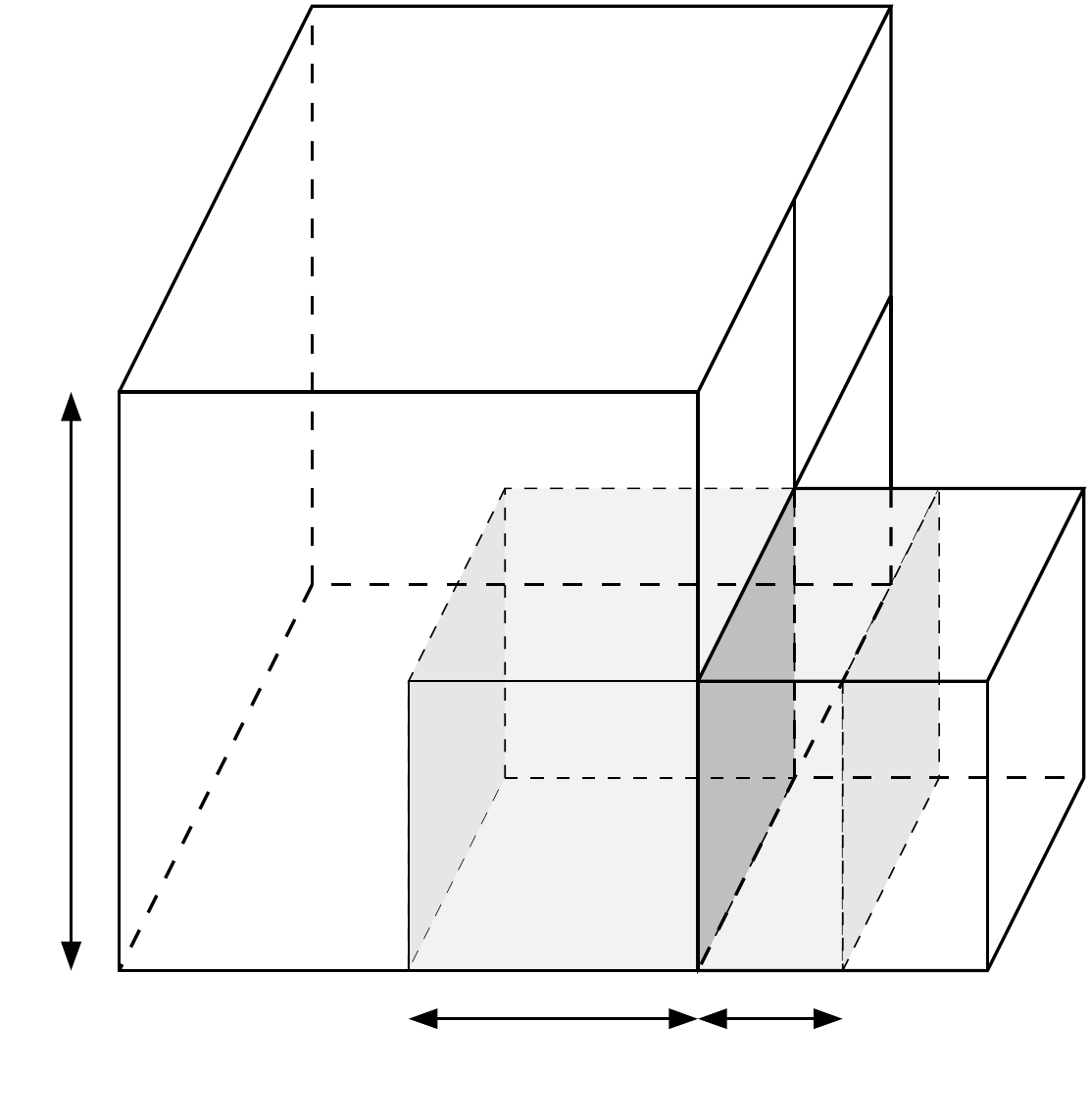
 \caption{ Left: $u_h(\by)$ is defined  by a linear interpolation based on the fan triangulation with the center in $\bx_V$,
i.e. interpolation of $u_h(\bx_V)$, $u_h(\bx_1)$,$u_h(\bx_{10})$ in this example.
Right:
The shaded region is the  control volume $V'$ associated with  face $F$ shared by cells of different sizes.\label{controlvol_adv}
\label{fig:advection}}
\end{figure}

First, we describe how the advection and diffusion terms are treated in the interior of the computational domain.
Several authors, e.g., \cite{Xiu2001semi,Losasso:04}, adopted semi-Lagrangian method to handle the time derivative and the inertia terms  in finite difference discretizations of the momentum equations on the octree meshes.
In \cite{Olshanskii2013231} we found that semi-Lagrangian method on octree meshes can be either excessively diffusive or prone to instabilities for flows passing submerged objects. As an alternative, we consider a  higher order upwind finite volume scheme on the graded octree meshes, which is both stable and accurate. Further details and the verification of the formal accuracy order of method can be found in thesis \cite{TerekhovDisser}. For the completeness of the presentation we describe the method below.

 In several places further in the text we need an  approximation of  the grid velocity function $\ba$ in an arbitrary point of the computational domain. For a given point $\by$ in the computational domain  we evaluate $\ba(\by)$ as follows. Assume $\by$ belongs to a cell $V$ and we are interested in interpolating the $x$-component of velocity to $\by$, i.e. $a_x(\by)$. Consider a plane $\cP$ such that $\by\in \cP$ and $\cP$ is orthogonal to the $Ox$ axis. Let $\bx_V\in \cP$ be the orthogonal projection of the center of $V$ on $\cP$ and $\bx_k$, $k=1,\dots,m$, $m\le12$, are the projections of centers of all cells sharing a face with $V$. The values $a_x(\bx_V)$ and $a_x(\bx_k)$ can be defined by a linear interpolation of the velocity values at nodes where $a_x$ is collocated.
Once    $a_x(\bx_V)$ and $a_x(\bx_k)$, $k=1,\dots,m$, are computed, we  consider the triangle fan based on $\bx_V$ and $\bx_k$, $k=1,\dots,m$, as shown in Figure~\ref{fig:advection} (left).
Now $a_x(\by)$ is defined  by a linear interpolation between the values of $a_x$ in the vertices of  the  triangle, which contains $\by$. {\color{black} The proposed interpolation procedure is faster and produces smaller stencil compared to a straightforward least squares fitting of a polynomial to velocity values in a set of nodes.}

For the incompressible fluid we treat the inertia terms in the `conservative' form $\bu\cdot\nabla\bu=\mathbf{div}\,(\bu\otimes\bu)$, where the
vector $\mathbf{div}\,$ operator applies row-wise.
Equation \eqref{step1} of the splitting method linearizes the nonlinear terms, so that we need to approximate $\mathbf{div}\,(\bu\otimes\ba)$ for
a given nodal velocity $\ba=(a_x,a_y,a_z)^T$ and unknown nodal velocity $\bu=(u,v,w)^T$.  Below we discuss the FV discretization of $\Div(u\ba)$. Other two components of  $\mathbf{div}\,(\bu\otimes\ba)$ are treated similarly.


Consider the velocity component $u$ at the $x$-node $\bx_F$, which is the barycenter of the face $F$.
If $F$ is shared by the  cells of different sizes, we define the control volume $V'$ as shown in Figure~\ref{controlvol_adv} (right).
If $F$ is shared by the cells of the same size, then $V'$ is defined in the obvious way by merging two half-cells.
Let $\F(V')$ denote the set of all faces for $V'$. We have
\begin{equation}\label{defconv}
\Div(u\ba)(\bx_F)\approx|V'|^{-1}\sum_{F'\in{\cal F}(V')}|F'|(\ba \cdot \bn) (\bx_{F'}) u(\bx_{F'}).
\end{equation}
We need to define advective fluxes at the barycenters $\bx_{F'}$ of faces  $F'\in{\cal F}(V')$.

\begin{figure}
 \centering
  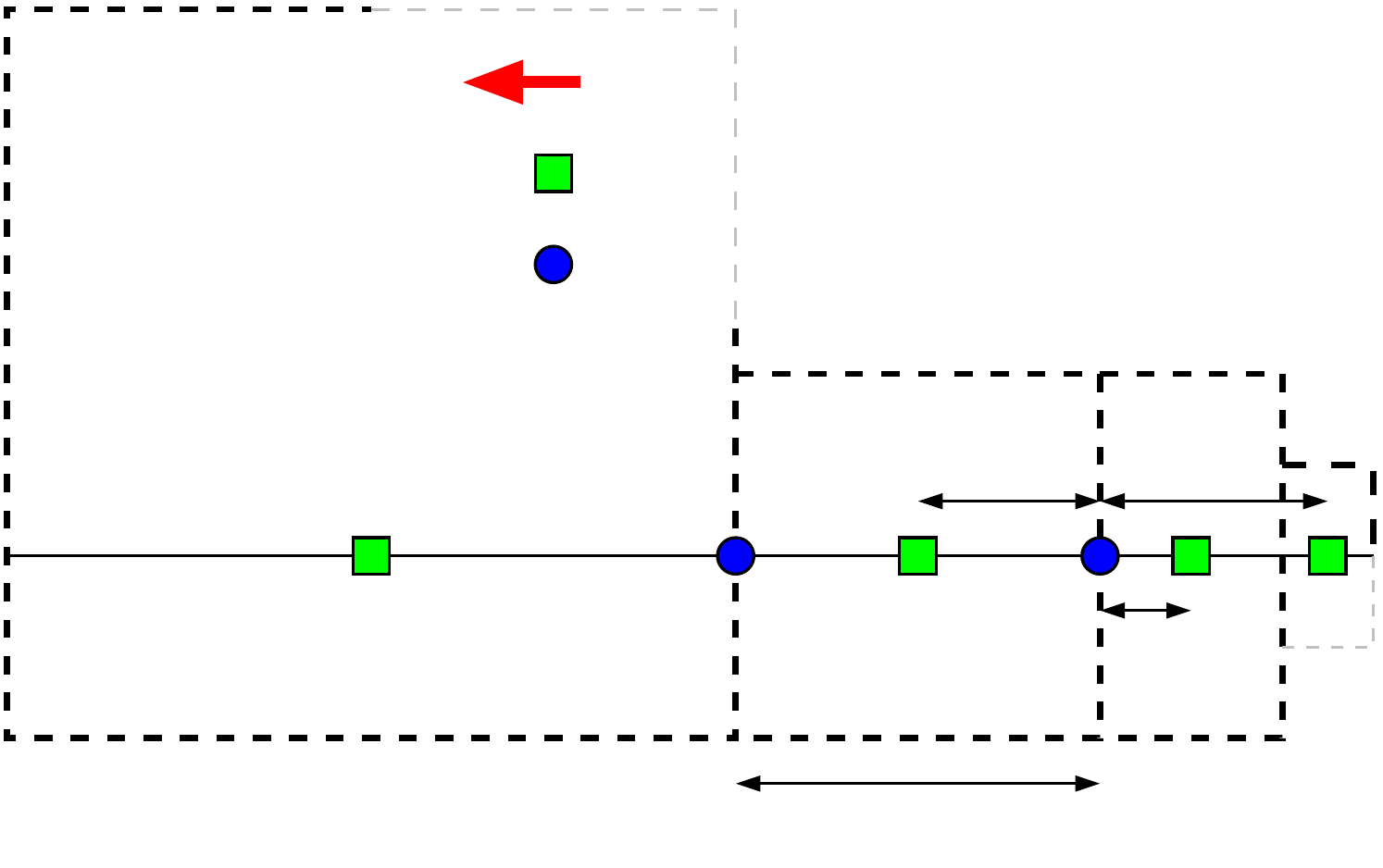\qquad
   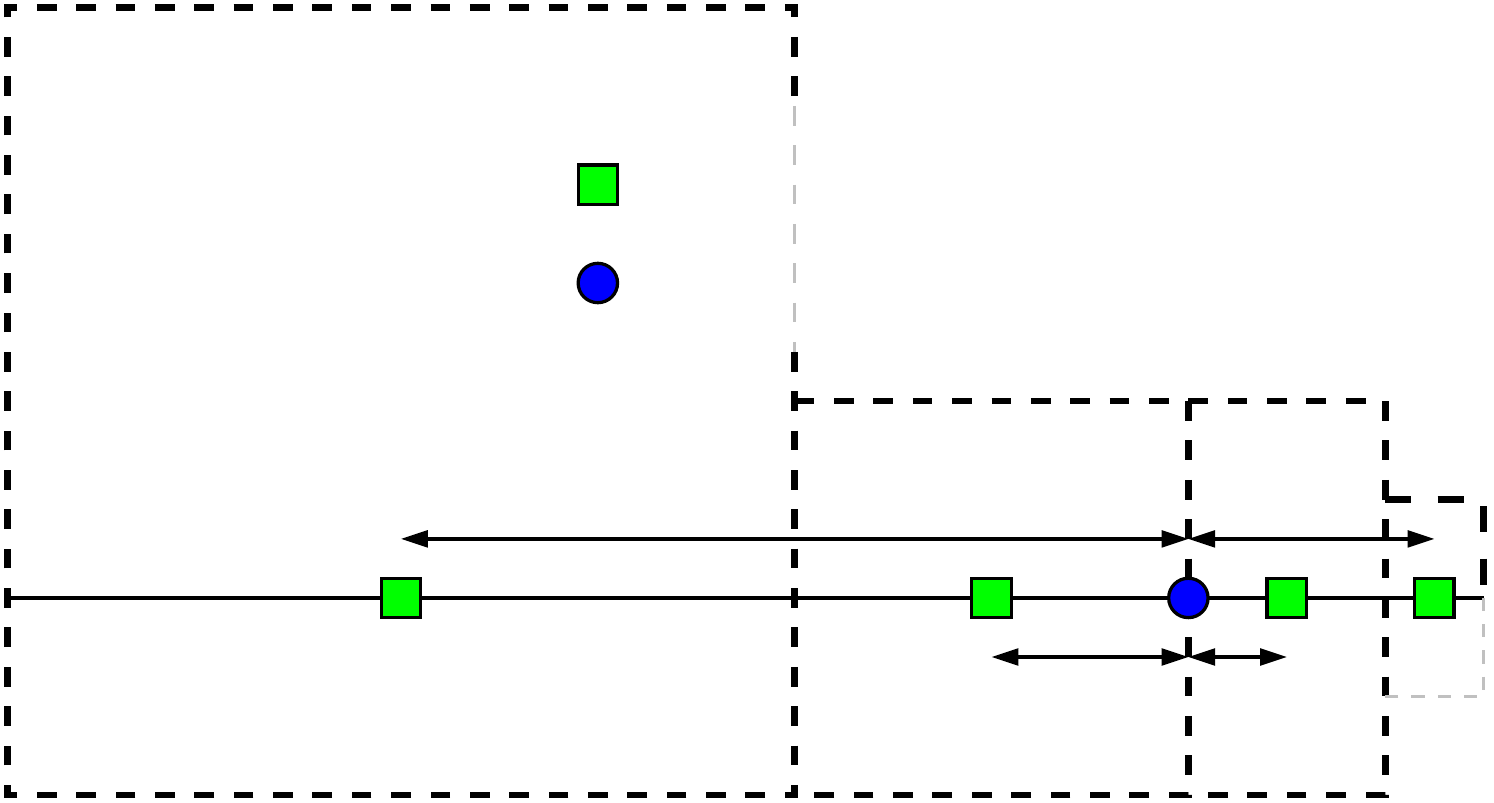
 \caption{Left: Reference points for the  upwind approximation of advection.
This illustration is for the derivative tangential to a face $F$, where the velocity degree of freedom is located.
Right: Reference points for the diffusion flux approximation\label{fig:diffusion}}
\end{figure}

First, we discuss the approximation of the advective flux at $F'\in{\cal F}(V')$ orthogonal to $F$.  Consider $F'$ orthogonal to $Oy$ so that $\ba \cdot \bn= a_y$.
If two cells sharing $F$ have the same size, then $(\ba \cdot \bn) (\bx_{F'})$ is the simple averaging of $a_y$ values from the two neighboring nodes. Otherwise   $a_y(\bx_{F'})$ is computed by the interpolation procedure described above.
To define $u(\bx_{F'})$, we take four `reference' points ($\bx_{-1}$, $\bx_{1}$, $\bx_{2}$, {$\bx_0:=\bx_F$}) as shown in Figure~\ref{fig:diffusion} (left).
Note that   $\bx_{-1}$, $\bx_{1}$, and $\bx_{2}$ are not necessarily grid nodes. Values $u_{-1}$, $u_{1}$, and $u_{2}$  in these nodes are then defined based on the following interpolation procedure.

If the reference point belongs to a cell \textit{smaller}  than the cell of $\bx_0$ (points $\bx_{1}$ and $\bx_{2}$ in the figure), then the linear interpolation between  the \textit{two} barycenters of adjunct faces is used.
If  the node belongs to a cell \textit{larger}  than the cell of $\bx_0$ (point $\bx_{-1}$  in the figure), then one apply the same interpolation procedure as we used above to define the values of $\ba$. The only difference is that instead of the linear interpolation using the fan triangulation for $\bx_V$ we use the weighted least-square method to fit the velocity values  $u(\bx_V)$ and $u(\bx_k)$ by the second order polynomial $Q_2$, and further set $u(\bx_{-1}):=Q_2(\bx_{-1})$.

If $a_y(\bx_F)>0$, the $u$-values in reference points $\bx_{-1}$, $\bx_{0}$, $\bx_{1}$ are used to approximate the flux. Otherwise,
the $u$-values in the reference points $\bx_{0}$, $\bx_{1}$, $\bx_{2}$
are needed. Assume $a_y(\bx_F)<0$, we set
\begin{multline}\label{ur}
 u(\bx_{F'})=D^{-1}\left[u_0(hH^2-h^2H)+u_1(rH^2+r^2H)-u_2(hr^2+h^2r)\right. \\
\left. +\lambda \Delta x^2(u_0(H-h)-u_1(H+r)+u_2(r+h))\right],
\end{multline}
where $D=(r+h)(H-r)(H+r)$. A family of formally second order upwind discretization is parameterized by  $\lambda\in\mathbb{R}$.
We found that $\lambda=0$
(defining the QUICK scheme \cite{Leonard_QUICK} on uniform meshes) produces the most accurate results on octree meshes and we use this value for numerical experiments.

Now, consider  the approximation of the advective flux at $F'\in{\cal F}(V')$ parallel to $F$,  hence $\ba \cdot \bn= a_x$.
After prescribing  $a_x(\bx_{F'})$  value with the help of the linear interpolation at the corresponding faces of the control volume, we define $u(\bx_{F'})$ using \eqref{ur}.
 The only differences with the treatment of the face $F'$ orthogonal to $Oy$ are the following:
$a_x$ is defined in $\bx$ (no interpolation required), and the reference points $\bx_{-1}$, $\bx_{1}$, $\bx_{2}$ are always lying on cells $x$-faces (although not necessarily in the centers and one has to do the interpolation).

Next, we explain how the discretization of viscous terms is computed.
Consider a node $\bx$ holding the velocity component $u$ and lying on a face $F$ and  define a cubic control volume $V'$ such that  $\bx$ is the center of $V'$ and $F$ is a middle cross section of $V'$. Note that the control volumes for $x$-nodes do not overlap, but for locally refined mesh they do not necessarily cover the whole bulk domain. Hence the dicretization of the viscous terms is a finite difference method, rather than a finite volume method.
We have
\begin{equation}\label{defdelta}
(\Delta_h u)(\bx)\approx |V'|^{-1}\sum_{F'\in{\cal F}(V')}|F'|(\nabla_h u \cdot \bn) (\by_{F'}).
\end{equation}
To approximate the diffusion flux at the center $\by_{F'}$
of $F'\in{\cal F}(V')$, we take four reference points ($\bx_{-1}$, $\bx_0$,  $\bx_{1}$, $\bx_{2}$)
as shown in Figure~\ref{fig:diffusion} ({\color{black}right}). Velocity values $u_{-1}$, $u_0$,  $u_{1}$, and $u_{2}$ are
assigned to reference points same way as for the advective terms described above.
Using the notation from Figure~\ref{fig:diffusion}, the formal third order approximation of the diffusion flux density $(\nabla u \cdot \bn)$ can be written out as
\begin{equation}\label{difflux}
\begin{aligned}
(\nabla u \cdot \bn) \approx
D^{-1}\big[&
  (h^2 H^3+h^3 R^2-H^3 R^2+h^2 R^3-H^2 R^3-h^3 H^2)   u_{0}\\
&+(H^3 R^2+r^3 R^2+H^2 R^3-r^2 R^3-H^3 r^2-H^2 r^3)   u_1\\
&+(h^3 r^2+h^2 r^3-h^3 R^2-r^3 R^2-h^2 R^3+r^2 R^3)  u_{-1}\\
&+(h^3H^2-h^2 H^3-h^3 r^2+H^3 r^2-h^2 r^3+H^2 r^3)  u_2
\big],
\end{aligned}
\end{equation}
with $D=(H-h) (h+r) (H+r) (h+R) (H+R) (R-r)$. If the reference point in $\bx_2$ is not available, we use the point $\bx_{-2}$.

To enforce incompressibility condition, we approximate $\Div\bu$ in the center $\bx_V$ of a  grid cell $V$. 
 We define the grid divergence operator by
\begin{equation}\label{defdiv}
(\Div_h \bu_h)(\bx_V)=|V|^{-1}\sum_{F\in{\cal F}(V)}|F|(\bu_h \cdot \bn) (\bx_F).
\end{equation}
Thanks to the staggered location of velocity nodes, the fluxes  $(\bu_h \cdot \bn) (\bx_F)$ are well-defined.

One way to introduce the discrete gradient is to define it as the adjoint of the discrete divergence.
We found that an approximation of $\nabla_h$  based on the formal Taylor expansions gives more accurate results.
For every internal  face  we assign the corresponding component of  $\nabla_h p$ as follows.
Since the octree mesh is graded, 
there can be only two geometric cases. If a face is shared by two equal-size cells, then the central difference approximation is used. Otherwise, for the approximation of $p_x$ at the face center node $\by$ one
considers the centers of five surrounding cells $\bx_1,\dots,\bx_5$
and expand the pressure value $p(\bx_i)$ with respect to $p(\by)$:
$$ p(\bx_i) = p(\by) + \nabla p(\by) \cdot (\bx_i - \by) + O (|\bx_i - \by|^2).$$
Neglecting the second-order terms, we obtain the following over-determined  system:
\begin{equation}\label{star}
\left(
\begin{array}{cccc}
1 & -{\Delta}/{2} & {\Delta}/{4} &{\Delta}/{4} \\[+1pt]
1 & {\Delta}/{4}  &  0 & 0 \\[+1pt]
1 & {\Delta}/{4}  & {\Delta}/{2} & 0 \\[+1pt]
1 & {\Delta}/{4}  &  0           & {\Delta}/{2} \\[+1pt]
1 &{\Delta}/{4}  & {\Delta}/{2} & {\Delta}/{2}
\end{array}
\right)
\left(
\begin{array}{c} p(\by) \\[+3pt]
p_x (\by) \\[+3pt]
p_y (\by) \\[+3pt]
p_z (\by)
\end{array}
\right) =
\left(
\begin{array}{c} p(\bx_1) \\[+1pt]
                 p(\bx_2) \\[+1pt]
                 p(\bx_3) \\[+1pt]
                 p(\bx_4) \\[+1pt]
                 p(\bx_5)
\end{array}
\right),
\end{equation}
where $\Delta \equiv \Delta x$.
The least squares solution of \eqref{star} gives the stencil for the $x$-component of the gradient:
\begin{equation}\label{defp}
p_x(\by) \approx \Frac{1}{3\Delta} (p_2+p_3+p_4+p_5-4p_1).
\end{equation}
The superposition of the discrete gradient and divergence operators generally leads to the non-symmetric matrix
for the pressure problem. However, the corresponding linear algebraic systems are solved efficiently
by a  Krylov subspace method with {\color{black}a two-parameter threshold ILU preconditioner~\cite{kaporin1998high,konshin2015ilu}}. {\color{black} We note that in general non-symmetric FV approximations of diffusion equations may lead to the lack of coercivity and hence to  stability issues, cf.~\cite{droniou2014finite}. Nevertheless, the previous studies, e.g., \cite{Losasso:04,Popinet2003572,Olshanskii2013231,NikitinVassilevski:08}, show that using the present non-symmetric approximations of the pressure Poisson equation does not disrupt  the stability of projection methods.}

It was noted in \cite{Olshanskii2013231} for octree staggered  grids, {\color{black} that} the discrete Helmholtz decomposition, which  essentially constitutes the projection step of the splitting scheme,  is unstable due to oscillatory spurious \textit{velocity} modes tailored to course-to-fine grid interfaces. If the viscosity  is sufficiently large, then  such modes are suppressed, otherwise they propagate and destroy the accuracy of numerical solution.
Following that paper we apply a technique, which eliminates  the spurious modes and improves the accuracy of numerical solution significantly.

The constructed spacial discretization is hybrid: a finite volume method was used to handle the incompressibility constraint and inertia terms, while a finite difference method was applied to diffusion terms and pressure gradient. {\color{black}To solve the velocity equation on each time step, we use BiCGStab(2)~\cite{sleijpen1993bicgstab} iteration with a two-parameter threshold ILU preconditioner~\cite{kaporin1998high,konshin2015ilu}. This combination of the Krylov subspace method and the preconditioner resulted in a robust and efficient solver.}

\subsection{Boundary conditions and curvilinear boundaries} \label{sec:curv}

The discretization method in section~\ref{sec:spartial} assumes that velocity values in all nodes forming flux stencils {\color{black} are given}.
When all the cubic volumes in the stencil are internal, then all corresponding velocity values are treated as active  degrees of freedom.  Close-to-boundary cells require special treatment. Below we introduce such a treatment when a curvilinear boundary is immersed in the background octree mesh.

\begin{figure}[h!]
 \centering
  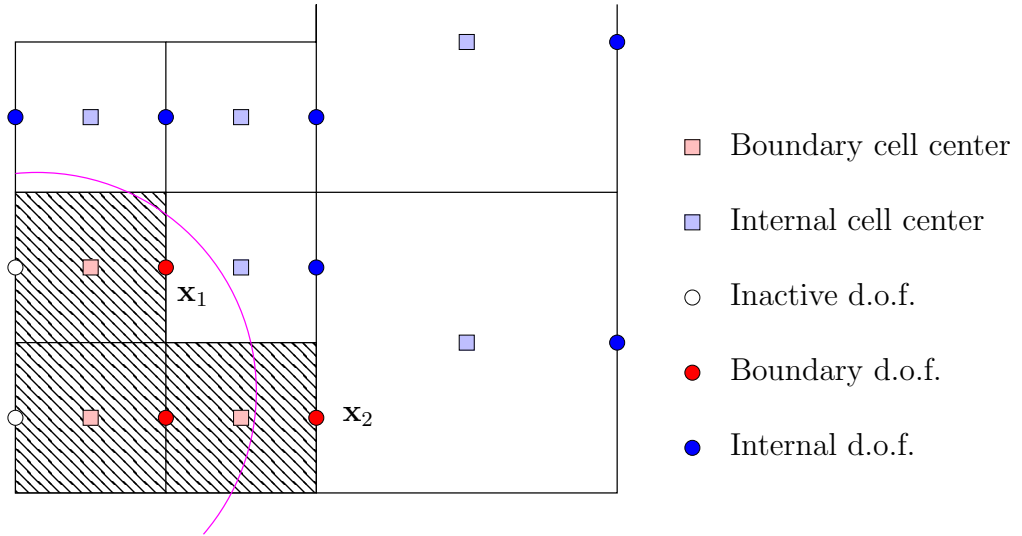
 \caption{Internal (blue), boundary (red) and inactive  (white) nodes  near the curvilinear boundary.}
 \label{fig:fict_dof_layouts}
\end{figure}

For the computational purposes, the static boundary is defined with the help of a signed distance function $\varphi_{s}$. {\color{black}We assume that the static boundary consists of several smooth components.  Each component is described by its own  $\varphi_{s}$ (domains of definition of the level set functions may overlap).}
This is  similar to the description of the free surface, but $\varphi_s$ is defined by the domain geometry and does not vary in time.
We assume that $\varphi_s < 0$ in  the fluid domain $\Omega$,
and $\varphi_s > 0$ in the exterior, so the boundary is given as the zero isosurface of function $\varphi_s$.
Denote by $\T_h$ the background octree mesh, the collection of cubic volumes forming the tessellation of the bulk computational domain. For each $V\in\T_h$, $\bp_V$ denotes the barycenter of $V$. We divide $\T_h$  into the sets of internal, boundary and external cells:
\[
\begin{split}
\T_{int}&:=\{V\in\T_h\,:\,\varphi_s(\bp_V)\le -h_{thr}\},\\ \T_{bdr}&:=\{V\in\T_h\,:\,\exists\,K\in \T_{int},~ s.t.~\overline{V}\cap \overline{K}\neq\emptyset\}, \qquad \T_{ext}:=\T_h\setminus(\T_{int}\cup\T_{bdr}),
\end{split}
\]
where  $h_{thr}=h_V/10$  is a threshold parameter. Based on this splitting we also divide all velocity nodes on $\T_h$ into three groups. Denote by $\N_h$ the collection of all velocity nodes from the bulk computational mesh. The nodes on the boundary of the bulk domain are not active. Any other node $\bx\in\N_h$ has exactly two cells $V^1_\bx$ and $V^2_\bx$ such that $\bx\in \bar V^1_\bx\cap \bar V^2_\bx$.
Now we divide $\N_h$  into the sets of internal, boundary and external nodes:
\[
\begin{split}
\N_{int}&:=\{\bx\in\N_h\,:\,V^1_\bx\in\T_{int} ~\text{and}~ V^2_\bx\in \T_{int}\},\\ \N_{bdr}&:=\{\bx\in\N_h\,:\,V^1_\bx\in\T_{bdr} ~\text{or}~ V^2_\bx\in \T_{bdr}\}, \qquad \N_{ext}:=\N_h\setminus(\N_{int}\cup\N_{bdr}).
\end{split}
\]
The velocity degrees of freedom are assigned to the internal nodes and boundary nodes, i.e. those from $\N_{int}\cup\N_{bdr}$.
There is a difference, however, how the method works for the nodes from $\N_{int}$ and $\N_{bdr}$: For each node from  $\N_{int}$ we have a set of algebraic equations derived in the previous section, while  each node from  $\N_{bdr}$ receives an auxiliary equation  based on boundary conditions.
The nodes from $\N_{ext}$ are not active.
 This subdivision of velocity nodes into three groups based on the position of the immersed boundary is illustrated in Figure \ref{fig:fict_dof_layouts} (the figure shows a 2D mesh and only nodes for the horizontal velocity component).

Now we  derive equations for the nodes from $\N_{bdr}$.  For the Dirichlet boundary condition
$\bu=\mathbf{u}_b$ on the immersed boundary, this is done componentwise as follows.
For each boundary node $\bx$ either interpolation or extrapolation procedure is performed depending on the sign of $\varphi_s(\bx)$. 

\begin{figure}[h!]
 \centering
 \includegraphics[width=\linewidth]{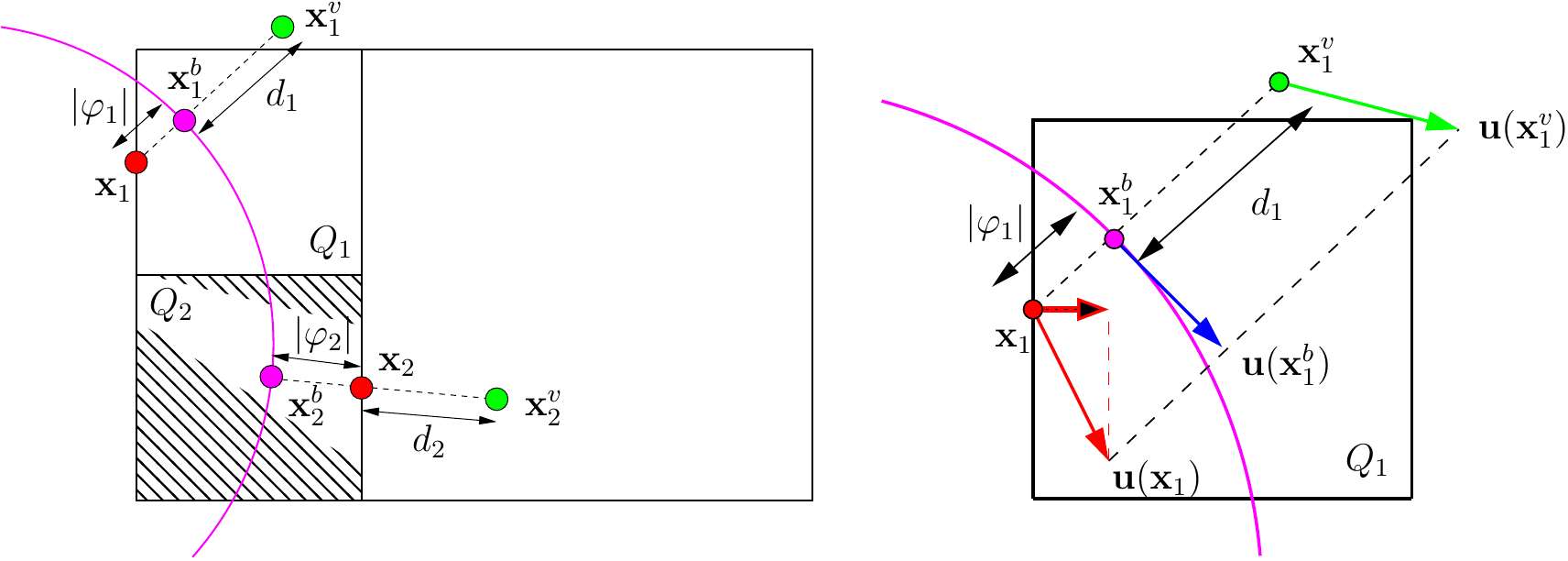}
 \caption{Left: Extrapolation of velocity values at the boundary node $\bx_1$ and interapolation at the boundary node $\bx_2$ near the curvilinear boundary; Right: Extrapolation of the velocity values at the boundary node $\bx_1$ for the free-slip boundary condition. \label{fig:fict_dof_int_ext} }

\end{figure}

For $\varphi_1 =\varphi_s(\bx_1) > 0$ (the node $\bx_1$ is outside the domain $\Omega$) we apply  extrapolation, cf. Figure~\ref{fig:fict_dof_int_ext} (left):
\begin{equation}
 \label{eq:fict_extra}
 u(\bx_1) = \frac{d_1 + \varphi_1}{d_1} u_b(\bx^b_1) - \frac{\varphi_1}{d_1} \,u(\bx^v_1),
\end{equation}
where $\bx^b_1$ is the closest boundary point to $\bx_1$,
$d_1 = \mathrm{max} (\varphi_1, h_{V_1})$ is an outstep to the internal domain,
and $\bx^v_1$ is a virtual node belonging to line passing through $\bx_1$ and $\bx_1^b$, and $|\bx_1-\bx_1^v| = d_1$.
The velocity value is  interpolated to $\bx^v_1$ from internal velocity degrees of freedom.

For $\varphi_2 =\varphi_s(\bx_2) <0$ (the node $\bx_2$  is inside the domain $\Omega$) we set
\begin{equation}
 \label{eq:fict_interp}
 u(\bx_2) =  \frac{d_2}{d_2 - \varphi_2}\,u_b(\bx^b_2) - \frac{\varphi_2}{d_2 - \varphi_2}\,u(\bx^v_2),
\end{equation}
where $\bx^b_2$ is the closest boundary point to $\bx_2$,
$d_2 = \mathrm{max} (-\varphi_2, h_{V_2})$ is an outstep to the external domain
and $\bx^v_2$ is a virtual node belonging to line $(\bx_2,\bx_2^b)$, $|\bx_2-\bx_2^v| = d_2$.
Again the velocity value is  interpolated to $\bx^v_2$ from internal velocity degrees of freedom.

For the free-slip boundary condition we use the approach similar to the no-slip condition.
Consider the boundary node $\bx_1$ and the virtual point $\bx^v_1$ with all velocity components $\bu(\bx^v_1)$ interpolated in it, see Figure~\ref{fig:fict_dof_int_ext} (right).
First, we write down the set of equations for $\bx_1$ assuming for a moment that all three components of $\bu$ are defined in $\bx_1$.
Thus, we seek for $\bu(\bx_1)$ such that interpolated (or extrapolated)  boundary value $\bu(\bx^b_1)$
has the normal component  vanishing and tangential components equal to those in the internal virtual node. This yields the following equations
\begin{equation*} 
\left\{
\begin{split}
\bu(\bx^b_1)& = \frac{d_1}{\varphi_1 + d_1} \bu(\bx_1) + \frac{\varphi_1}{\varphi_1 + d_1} \bu(\bx^v_1),\\
\bu(\bx^b_1) \cdot \bn &= 0,\\
\bu(\bx^b_1) - (\bu(\bx^b_1) \cdot \bn) \, \bn &= \bu(\bx^v_1) - (\bu(\bx^v_1) \cdot \bn) \, \bn,
\end{split}\right.
\end{equation*}
where $\bn$ is the unit normal vector for the boundary in point $\bx^b_1$.

Substituting the first and the second equations in the third one, we get the equation for $\bu(\bx_1)$:
\begin{equation} \label{eq:free-slip}
\begin{split}
\bu(\bx_1) = \bu(\bx^v_1) - \frac{\varphi_1 + d_1}{\varphi_1} (\bu(\bx^v_1) \cdot \bn) \, \bn.
\end{split}
\end{equation}
The final  equation tailored to the node $\bx_1$ follows by extracting only one equality from \eqref{eq:free-slip}. This equality  corresponds to the component of $\bu$ located at $\bx_1$.


The boundary condition \eqref{eq:tension} on the free surface and $\Gamma_{out}$ is decomposed into the homogeneous Neumann  boundary condition for the auxiliary velocity in the convection-diffusion step \eqref{step1} and the homogeneous Dirichlet boundary condition for the pressure correction variable $q$ in \eqref{step3}. For the pressure Dirichlet  condition the missing values at the barycenters of boundary cells are recovered by the same technique as Dirichlet velocity values for  the boundary with the no-slip condition.
{\color{black} Therefore, the pressure field is known in all close-to-free-boundary cells and the pressure update \eqref{step3} is well defined in cells from $\T_{int}$, which
did not belong to $\T_{int}$ at time $t_n$. }
The Neumann velocity boundary condition is enforced in the same way as the slip-condition on $\Gamma_D$. Of course, no-penetration condition does not apply in this case.

{\color{black}
Note that boundary nodes  receive  velocity values \textit{implicitly} through equations \eqref{eq:fict_extra}, \eqref{eq:fict_interp}, or   \eqref{eq:free-slip}. These equations are added to the global system of algebraic equations. To obtain a complete system, we need to discretize the momentum and continuity equations in all cut cells. To this end, we first extend the density and viscosity coefficients  by the same constant values from the cut cells to the whole cubic cells. Next, we apply the ``full-cell'' expressions in \eqref{defconv}, \eqref{defdelta}, and \eqref{defdiv} to define discrete operators for the cut cells.  Due to the linear extrapolation of  boundary conditions, the resulting differences  approximate the required differential operators.
}

Poisson equation for the pressure correction $q$ of the projection step involves  degrees of freedom at pressure nodes,  i.e. at barycenters of cells from $\T_h$.
We solve for the pressure degrees of freedom only for cells from $\T_{int}$. Thus, the discrete gradient
is well defined at all velocity nodes from $\N_{int}$ with the help of the pressure correction values at $\T_{int}$.
The discrete gradient at the nodes from $\N_{bdr}$ for $\Gamma_2$ is also well defined
with the help of internal  degrees of freedom and zero Dirichlet  values for the pressure correction in the free-boundary  cells.
To assign the gradient of the pressure correction to the nodes from $\N_{bdr}$ for $\Gamma_1$, we proceed as follows: From \eqref{step2a} we
get $\bu^{n+1} =  \widetilde {\bu^{n+1}}+ \triangle t_n/\alpha\nabla q$. The variable $\widetilde {\bu^{n+1}}$ receives its values in  all nodes from $\N_{bdr}$ during the predictor step \eqref{step1} of the splitting algorithm. Further we substitute the equality in the corresponding equations from \eqref{eq:fict_extra}--\eqref{eq:free-slip} for $\bu^{n+1}$ and this yields
the equation for $\nabla q$ in the boundary nodes.
Further we build the pressure Laplace operator as the superposition of the  gradient \eqref{defp} and
divergence \eqref{defdiv} grid operators.

\begin{remark}\rm
{\color{black} A rigorous stability analysis of the hybrid method is an open question. We note that stability of the semi-discrete scheme from section~\ref{sec:split} (only discretization in time) for free-surface flows was studied in \cite{Nikitin2014}. The scheme was shown to conserve global momentum and angular momentum, and based on that an energy inequality was shown to hold. Thorough numerical studies of the stability and numerical dissipation of the method for the case of enclosed flows (no free boundary) and fitted boundary conditions (no curvilinear boundaries) was done in \cite{Olshanskii2013231}. In that paper, the method was shown  stable  for a vast range of flows (from laminar to developed turbulent); it was shown to have lower numerical diffusion compared to some alternative approaches on octree meshes. The numerical results of the present paper suggest that this stability property extends to flow problems with free boundaries and streamlined bodies.
}
\end{remark}

\section{Numerical experiments}
\label{sec:num}

Our first series of  numerical experiments aims to assess the stability of the presented method, its lower dissipation and ability to handle free surface evolution accurately. To this end, we consider several standard benchmark problems.

The first two benchmark tests  deal with laminar flows around a 3D cylinder of circular cross-section at Re=20 and varying Reynolds number. This problem  does not require a dynamic adaptation of the octree mesh. Our goal here is to check the accuracy of the scheme in a domain with curved boundary by comparing computed drag and lift coefficients with those found in the literature. These statistics are known to be sensitive to excessive numerical dissipation of a numerical method.

The lateral sloshing  tank benchmark   verifies the ability of the scheme to reproduce complex dynamics of fluid free surface.
The correctly recovered free surface evolution after the termination of  excitation forces  is another indicator of the scheme reliability and low numerical dissipation.
Dynamic mesh adaptation is very helpful in this problem.

After validation of the numerical scheme, we apply it to simulate  a water flow with surface waves around an oil platform rigidly mounted in the Kara sea offshore.

\subsection{Flow around  cylinder of circular cross-section}
The first  numerical test is the laminar 3D channel flow around a cylinder  of circular cross-section.
The problem was suggested as a benchmark by Sch\"{a}fer and Turek in \cite{ST96} and further studied in, e.g., \cite{John2002,BR2006,Rebholz2012}.

\begin{figure}  \centering
 \includegraphics[width=0.8\textwidth]{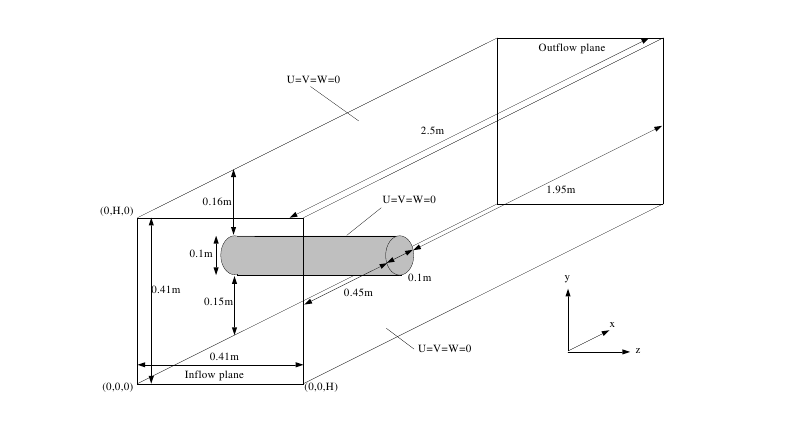}
 \caption{Computational domain for flow around cylinder of circular cross-section.\label{fig_cyl}}
\end{figure}

The flow domain  is shown in Figure \ref{fig_cyl}.
The no-slip and no-penetration boundary condition $\bu=0$ is prescribed on the channel walls and the cylinder surface.
For the outflow boundary conditions we put the normal component of the stress tensor equal zero on $\Gamma_{\rm out}$.
The parabolic velocity profile is set on the inflow boundary:
\[\bu = (0, 0, 16\widetilde{U}xy(H-x)(H-y)/H^4)^T\quad\text{on}~\Gamma_{\rm in},
\]
with $H=0.41$ and a peak velocity $\widetilde{U}$. The Reynolds number, $Re= \nu^{-1}D\widetilde{U}$, is defined based on the cylinder width $D=0.1$.
The viscosity coefficient  $\nu$ is set to $10^{-3}$.
We consider two benchmark tests from \cite{ST96}:
\begin{itemize}
\item Problem Z1: Steady flow with $Re=20$ ($\widetilde{U}=0.45$);
\item Problem Z3: Unsteady flow with  varying Reynolds number for $\widetilde{U}=2.25\sin(\pi t / 8)$.
\end{itemize}
The initial condition for both problems is $\bu=0$ for $t=0$.

The following statistics are of interest:
\begin{itemize}
\item The  difference $\Delta p = p(\bx_2) - p(\bx_1)$ between the pressure  values in points $\bx_{1} = \{0.2, 0.205,0.55\}$ and $\bx_{2} = \{0.2,0.205,0.45\}$.
\item The drag coefficient given by an integral over the surface of the cylinder $S$:
\begin{equation}\label{drag}
C_{\rm drag} = \frac{2}{D H  \widetilde{U}^2} \int_S \left( \nu \frac{\partial (\bu\cdot\mathbf{t})}{\partial \bn}n_x - p n_z \right) ds.
\end{equation}
Here $\bn=(n_x,n_y,n_z)^T$ is the normal vector to the cylinder surface pointing to $\Omega$ and $\mathbf{t}=(-n_z,0,n_x)^T$ is a tangent vector.
\item The lift coefficient  given by an integral over the surface of the cylinder:
\begin{equation}\label{lift}
C_{\rm lift} = - \frac{2}{D H  \widetilde{U}^2}  \int_S \left( \nu \frac{\partial (\bu\cdot\mathbf{t})}{\partial \bn}n_z + p n_x \right) ds.
\end{equation}
\end{itemize}

\begin{figure}  \centering
 \includegraphics[width=0.8\textwidth]{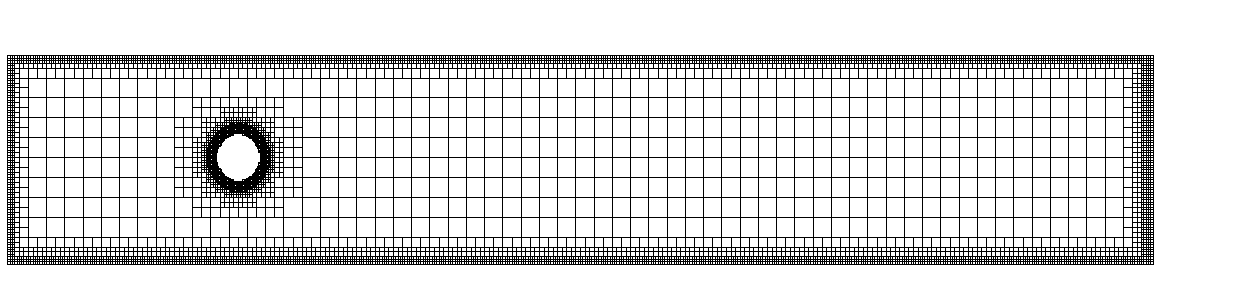}
 \caption{The cutaway of the grid at $y=0.205$ for $h_{max} = \ell/64$ and $h_{min} = \ell/1024$.\label{fig_grid}}
\end{figure}

The octree mesh is refined locally towards the channel walls (we set $h_{wall} = \ell/256$ except the coarsest mesh where $h_{wall} = \ell/128$,  $\ell=2.5m$ is the length of the computational domain) and the circular cylinder ($h_{\min}$ in this experiment denotes the mesh size near the cylinder).
The cutaway of the mesh with  $h_{min} = \ell/64$ and $h_{max} = \ell/1024$ is shown in Figure~\ref{fig_grid}.

To compute the drag and lift coefficients, we replace the surface integrals in \eqref{drag} and \eqref{lift} by
integration over the whole domain \cite{John2002,BR2006}: Assume $\bu=(u,v,w)^T$ and $p$ is the Navier-Stokes solution in a fixed domain $\Omega$, then applying the integration by parts one checks
the following identities:
\begin{equation} \label{draglift}
\begin{split}
C_{\rm drag} =\widetilde{C}\int_{\Omega}\left[ \left( \frac{\partial w}{\partial t} + \left(\bu \cdot \nabla\right)w\right) \varphi  + \nu\nabla w\cdot\nabla\varphi - p\partial_z \varphi\right]\,  \mathrm{d}\bx \\
C_{\rm lift} =\widetilde{C} \int_{\Omega} \left[ \left( \frac{\partial u}{\partial t} + \left(\bu \cdot \nabla\right)u\right) \varphi  + \nu\nabla u\cdot\nabla\varphi - p\partial_x \varphi\right]\,  \mathrm{d}\bx,
\end{split}
\end{equation}
$\widetilde{C}=\frac{2}{D H  \widetilde{U}^2}$, for any $\varphi\in H^1(\Omega)$ such that $\varphi |_{S} = 1$ and $\varphi |_{\partial \Omega / S} = 0$.
The accuracy of evaluation of \eqref{draglift} for a numerical solution depends on the regularity of $\varphi$.
In our numerical scheme $\varphi$ is defined in  pressure nodes as
the discrete harmonic function solving $\Div_h\nabla_h \varphi = 0$. The derivatives in  \eqref{draglift} are approximated with the second order of accuracy. Using the volume based  formulas \eqref{draglift} gives more accurate values of drag and lift coefficients compared to  \eqref{drag} and \eqref{lift}, if the Navier-Stokes solution is sufficiently smooth, see \cite{BR2006}.

\begin{table}\small
\begin{center}
\caption{\label{t_dof} The number of velocity and pressure d.o.f. for different meshes for problem Z1.}
\begin{tabular}{cc|rr}
\hline
$h_{min}$ & $h_{max}$ & $\bu$ d.o.f. & $p$ d.o.f.   \\
\hline
$\ell/128$ & $\ell/64 $  & 175126 & 65002     \\
$\ell/256$ & $\ell/64 $  & 855529 & 304395    \\
$\ell/512$ & $\ell/64 $  & 925177 & 338997    \\
$\ell/1024$ & $\ell/64$  & 1346577& 524983    \\
\hline
\end{tabular}
\end{center}
\end{table}

The numerical solutions to  problem Z1  were computed on a sequence of locally refined meshes, see Table~\ref{t_dof} for the information of the corresponding discrete space dimensions.
Note that we  refine the mesh sequence towards the cylinder and keep it coarser in the wake.
Such refinement is known to be  crucial for accurate computation of the statistics of interest, see, for example
\cite{BR2006,Olshanskii2013231}.

\begin{table}\small
\begin{center}
\caption{\label{turek_z1} Problem Z1: Convergence of drag, lift,  and pressure drop  to reference
 intervals.}
\begin{tabular}{cc|ccc}
\hline
$h_{min}$ & $h_{max}$ & $C_{\rm drag}$ & $C_{\rm lift}$ & $\Delta p$  \\
\hline
$\ell/128$ & $\ell/64$   &  3.07235 & -0.019821  & 0.13840   \\
$\ell/256$ & $\ell/64$   &  6.20151 &  0.00778   & 0.15961   \\
$\ell/512$ & $\ell/64$   &  6.15078 &  0.00962   & 0.16298   \\
$\ell/1024$& $\ell/64$   &  6.14193 &  0.00990   & 0.16636   \\
\hline
\multicolumn{2}{c}{Braack \& Richter} &  6.18533 & 0.009401 &   \\
\multicolumn{2}{c}{Sch\"{a}fer \& Turek} & 6.05--6.25 &  0.008--0.01 &  0.165--0.175 \\
\hline
\end{tabular}
\end{center}
\end{table}

The reference \cite{ST96} collects several DNS results based on various
finite element, finite volume discretizations of the Navier-Stokes equations and the Lattice Boltzmann method.
One can find there reference intervals where the statistics of interest should converge.
Using a  higher order finite element method and locally refined  adaptive meshes, more accurate
reference values of  $C_{\rm drag}$ and $C_{\rm lift}$ are found in~\cite{BR2006}
for  problem  Z1.
For a sequence of locally refined octree  meshes, Table~\ref{turek_z1} demonstrates the convergence
of computed drag and lift coefficients,  and pressure drop value to reference intervals.

\begin{table}\small
\begin{center}
\caption{\label{turek_z3} Problem Z3: Maximum drag, maximum/minimum lift,  and pressure drop (at $t=8$) and reference intervals. Time steps subject to $\Delta t_k = 5 h_{\min}/\max\limits_\bx |\bu(\bx,t_k)|$.}
\begin{tabular}{cc|cccc}
\hline
$h_{min}$ & $h_{max}$ & $\max C_{\rm drag}$ & $\max C_{\rm lift}$  & $\min C_{\rm lift}$ & $\Delta p (t=8)$  \\
\hline
$\ell/256$ & $\ell/128$  &3.74685        &0.00190        &-0.01474       &-0.09740\\
$\ell/512$ & $\ell/128$  &3.22627        &0.00329        &-0.01197       &-0.12083\\
$\ell/1024$& $\ell/128$  &3.13382        &0.00325        &-0.01011       &-0.11933\\
\hline
\multicolumn{2}{c}{Bayraktar \& Mierka \& Turek} & 3.29--3.33 & 0.0027--0.0033 & -0.010-- -0.012 &  \\
\multicolumn{2}{c}{Sch\"{a}fer \& Turek} & 3.2--3.3   &  0.002--0.004&  & -0.14-- -0.12\\
\hline
\end{tabular}
\end{center}
\end{table}

For problem Z3 less accurate reference data is available.
Table~\ref{turek_z3} summarizes the results computed by the present method and those available in the literature \cite{Bayraktar,ST96}.
The values of $\max C_{\rm drag}$, $\max C_{\rm lift}$ and $\min C_{\rm lift}$
are the maximum drag and maximum/minimum lift  coefficients over the whole time interval $t\in[0,8]$, the pressure
drop $\Delta p$ is computed at $t=8$. The most sensitive statistics are $C_{\rm lift}$ and $\Delta p$. Table~\ref{turek_z3} shows their convergence to the reference intervals. The value of the maximum drag coefficient on the finest mesh is slightly (3\%) less than the reference one.

\begin{figure}[!htb] \centering
\includegraphics[width=0.49\textwidth]{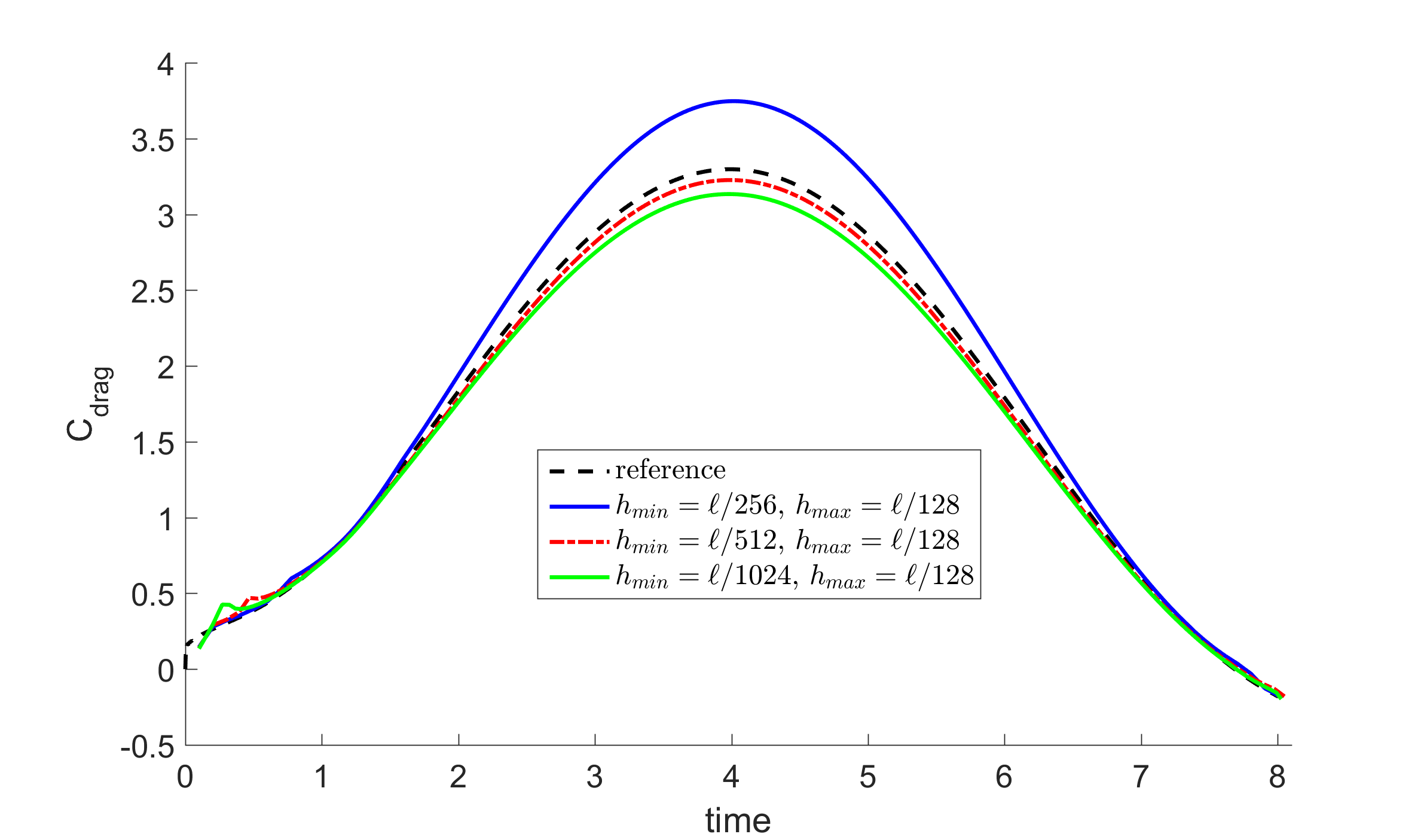}
\includegraphics[width=0.49\textwidth]{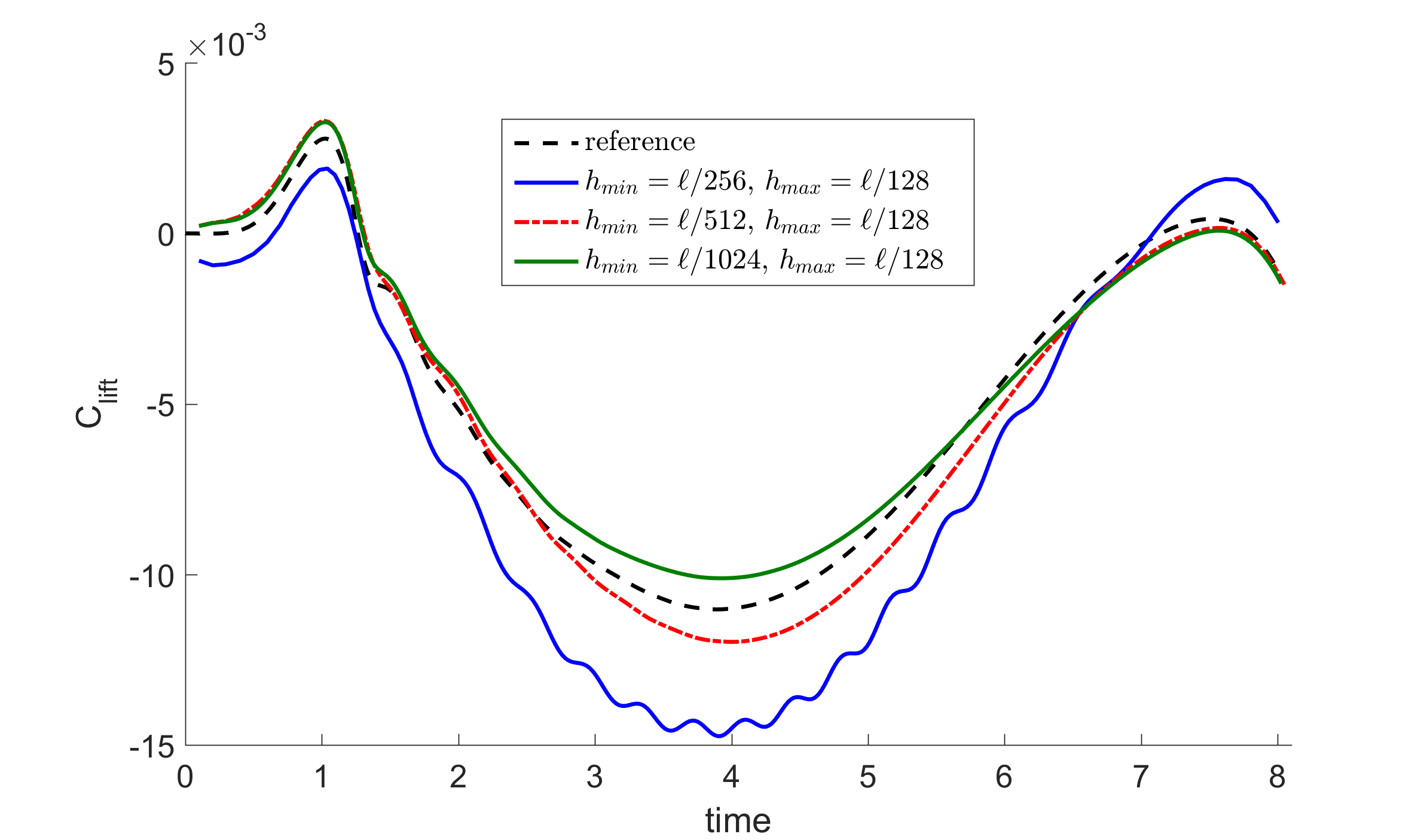}
\caption{Computed drag (left) and lift (right) coefficients dependence on time versus reference data from \cite{Bayraktar}.}
\label{fig:CdragCliftZ3}
\end{figure}

In Figure \ref{fig:CdragCliftZ3} we compare the computed curves $C_{\rm drag}(t)$, $C_{\rm lift}(t)$ with reference data from \cite{Bayraktar}. The computed coefficients fit the reference data reasonably well. {\color{black} For a better fitting, a stronger mesh refinement is needed which exceeds our computing capabilities.}

\subsection{Sloshing tank}

\begin{figure}[!htb] \centering
\includegraphics[width=0.49\textwidth]{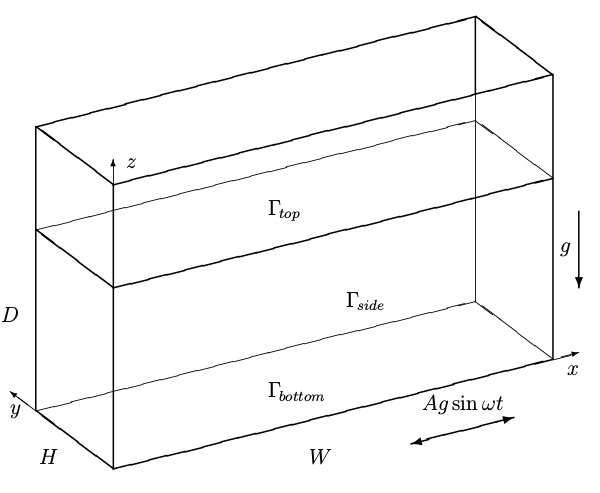}
\includegraphics[width=0.49\textwidth]{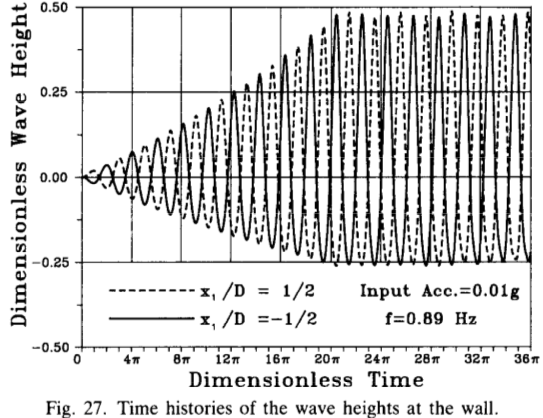}
\caption{Left: Problem setup for sloshing tank test. Right: Wave height  at
the mid-lines of two opposite tank walls from \cite{huerta88}}
\label{fig:domain_behr}
\end{figure}

The sloshing of fluid in a tank is a benchmark problem for numerical free surface flow solvers and a problem of independent interest, see, e.g.,   \cite{noh1964,huerta88,behr,frandsen2004sloshing,virella2008linear,chen2005time}.
The setup of the sloshing tank problem is given in \cite{huerta88, behr}.
A volume of water fills a rectangular  tank  as illustrated in Figure \ref{fig:domain_behr}.
The initial bulk dimensions are $W = 0.8 m$, $H = 0.1 m$  and $D = 0.3 m$.
The container walls  $\Gamma_{bottom}$ and $\Gamma_{side}$ impose slip and no-penetration conditions for the fluid.
The fluid is exposed to external forces: a constant gravitational acceleration of magnitude $g = 9.81ms^{-2}$ and  a sinusoidal horizontal excitation $Ag\;sin\;\omega t$ with $A = 0.01$ and $\omega = 2\pi f, f = 0.89 Hz$. The problem is non-dimensionalized  following \cite{huerta88}. The full set of
dimensional and non-dimensional parameters is summarized in Table \ref{tab:sloshing_params}.

\begin{table}
\caption{\label{tab:sloshing_params} Dimensional and non-dimensional parameters in the sloshing benchmark}
\begin{center}
\begin{tabular}{ l | c | c }
\hline
Value &Dimensional&Non-dimensional\\
\hline
Lengths&$D = 0.3\;m$&$\tilde{D}=1.0$\\
&$H = 0.1\;m$&$\tilde{H}=0.3333$\\
&$W = 0.8\;m$&$\tilde{W}=2.6667$\\
Frequency&$f=0.89s^{-1}$&$\tilde{f}=1.0$\\
Acceleration&$g=9.81ms^{-2}$&$\tilde{g}=1.0$\\
Viscosity&$\nu=1.0\times10^{-6}m^2s^{-1}$&$\tilde{\nu}=1.943\times10^{-6}$\\
\hline
\end{tabular}
\end{center}
\end{table}

The sloshing motion is initiated as soon as the horizontal excitation is applied.
After the initial ten periods the excitation is terminated.
The excitation frequency is designed to induce the first mode of wave motion in the $x$ direction, i.e.,
the motion with a wavelength approximately equal  the doubled  width of the tank $W$.
The time histories for the height of the wave at
the two opposite tank walls orthogonal to the $x$-axis are shown in Figure \ref{fig:domain_behr} (right).
These data were computed for the 2D setting of the problem in \cite{huerta88}. These results are believed to correspond well to physical observations \cite{noh1964}.

\begin{figure}[!htb] \centering
\includegraphics[width=0.49\textwidth]{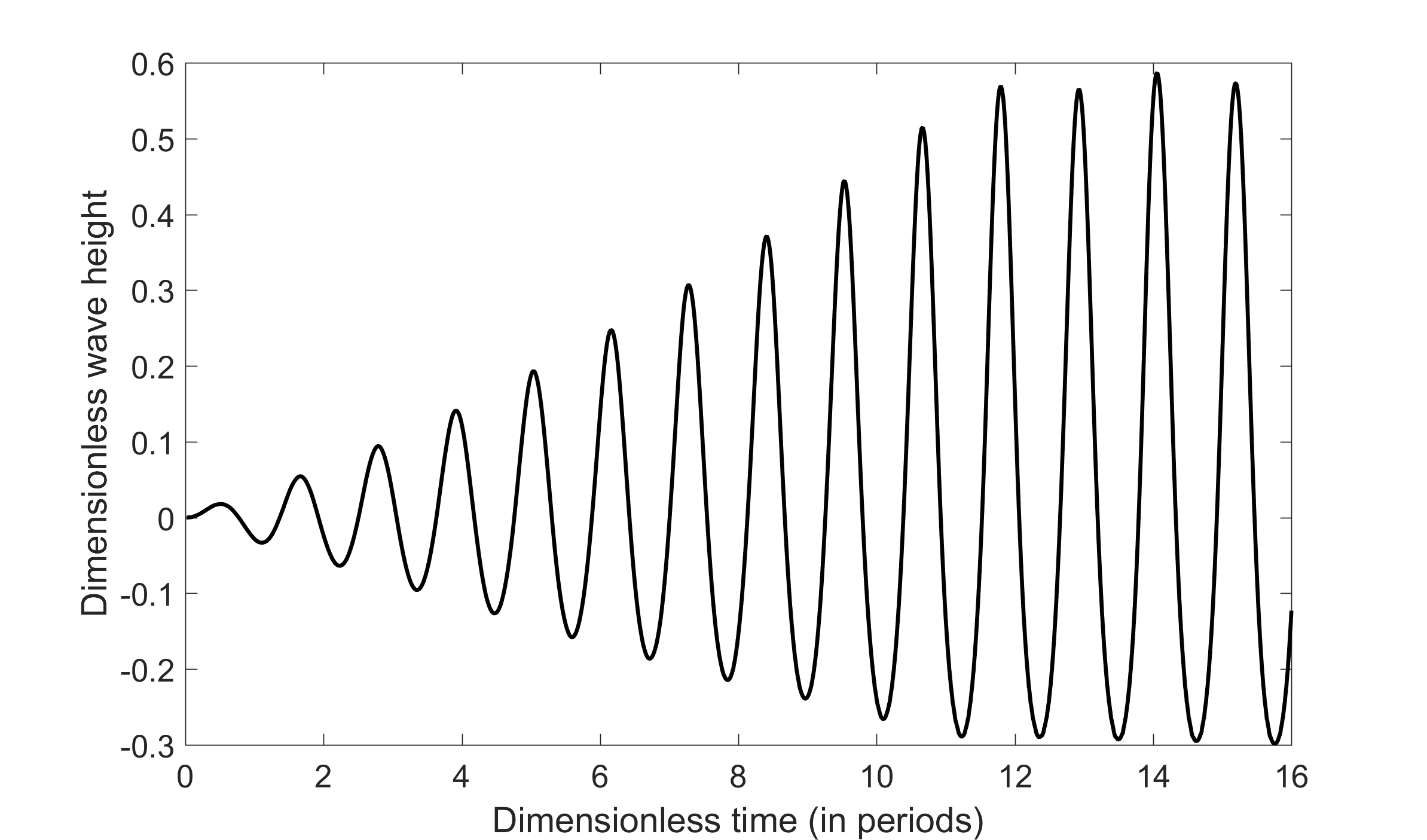}
\includegraphics[width=0.49\textwidth]{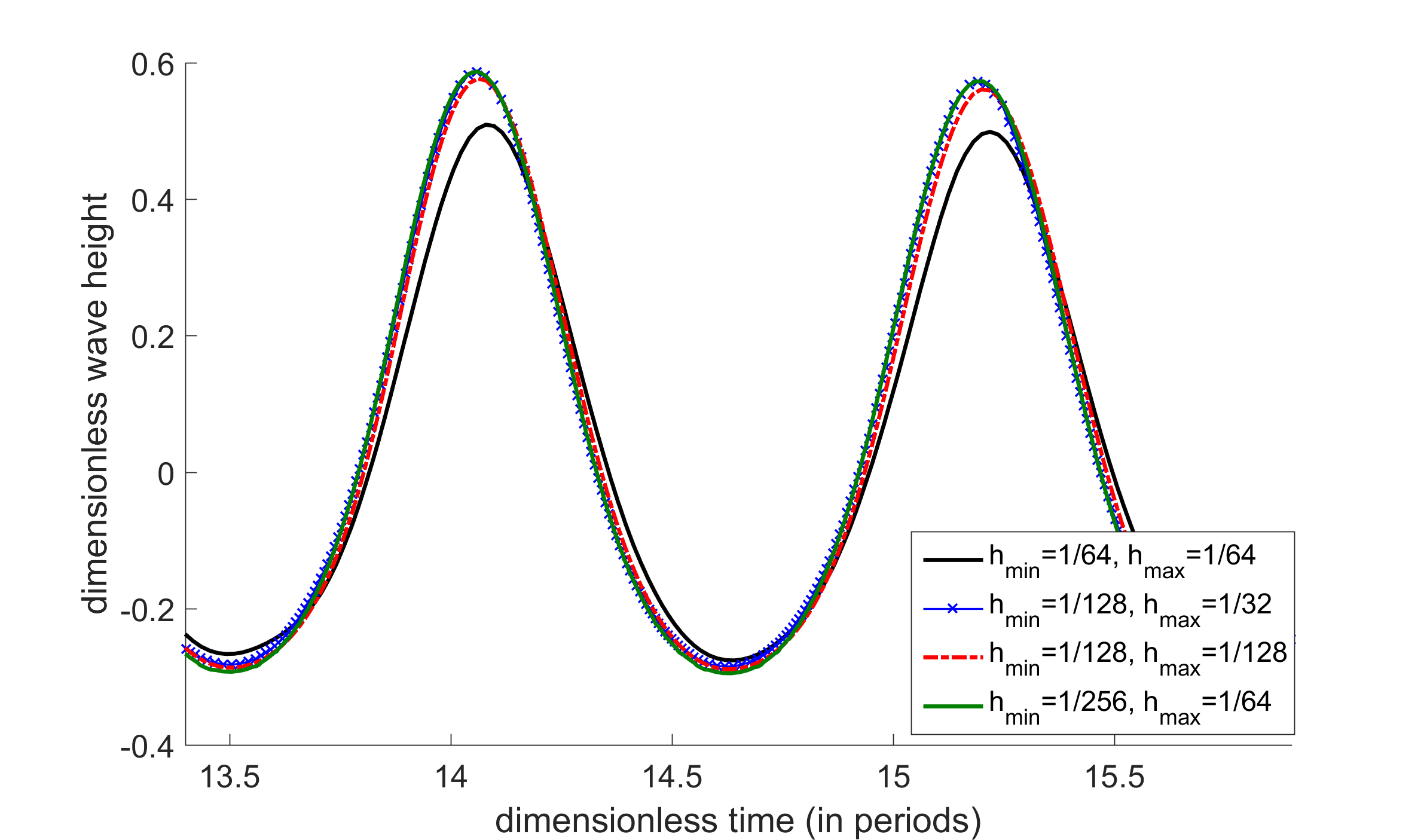}
\caption{Computed water level at the midline of the left wall ($x=-D/2$). Left plot shows the evolution computed with the
$h_{wall}=\ell/64$, $h_{\min}=\ell/256$. Right plot demonstrates the mesh-convergence of the results. }
\label{fig:sloshing_all}
\end{figure}

The octree FV method recovers correctly time dependence of the water level at the midline of the left wall ($x=-D/2$), see Figure \ref{fig:sloshing_all} (left).
For the first ten periods of excitation the measured wave height matches the heights
reported in \cite{huerta88} with the deviations less than  4\%. Numerical dissipation is low enough to avoid amplitude dumping after termination of the excitations
even on relatively coarse meshes.
The  mesh convergence of the free surface contact line evolution on the wall  at $x=-D/2$ is demonstrated in Figure \ref{fig:sloshing_all} (right).
The meshes are refined dynamically to the tank walls up to the meshsize $h_{wall}$ and to the free surface up to the meshsize $h_{\min}$, the coarsest cell size is fixed
$h_{\max} = \ell/8$, here $\ell=W$. {\color{black} At the ``Remeshing'' step of the splitting method we refine all cubic cells intersected by the zero level set of $\phi(t^{n+1})$ so that all these cells have the width $h_{\min}$. All other cells except boundary cells are marked for coarsening. The
coarsening is performed in such a way that the octree remains balanced (two neighbouring cells may differ
in size at most by a factor of two) and the maximum cell width in the fluid domain is $h_{\max}$.}
The following combinations of the mesh refinements were used: $h_{wall}=h_{\min}=\ell/64$, $h_{wall}=h_{\min}=\ell/128$, $h_{wall}=\ell/32$, $h_{\min}=\ell/128$, and $h_{wall}=\ell/64$, $h_{\min}=\ell/256$.
For this problem we use adaptive time step,   $\Delta t_k = \min\{0.0187,  h_{\min}/\max\limits_\bx |\bu(\bx,t_k)| \}$.

\begin{figure}
    \centering
        \includegraphics[width=0.53\textwidth,]{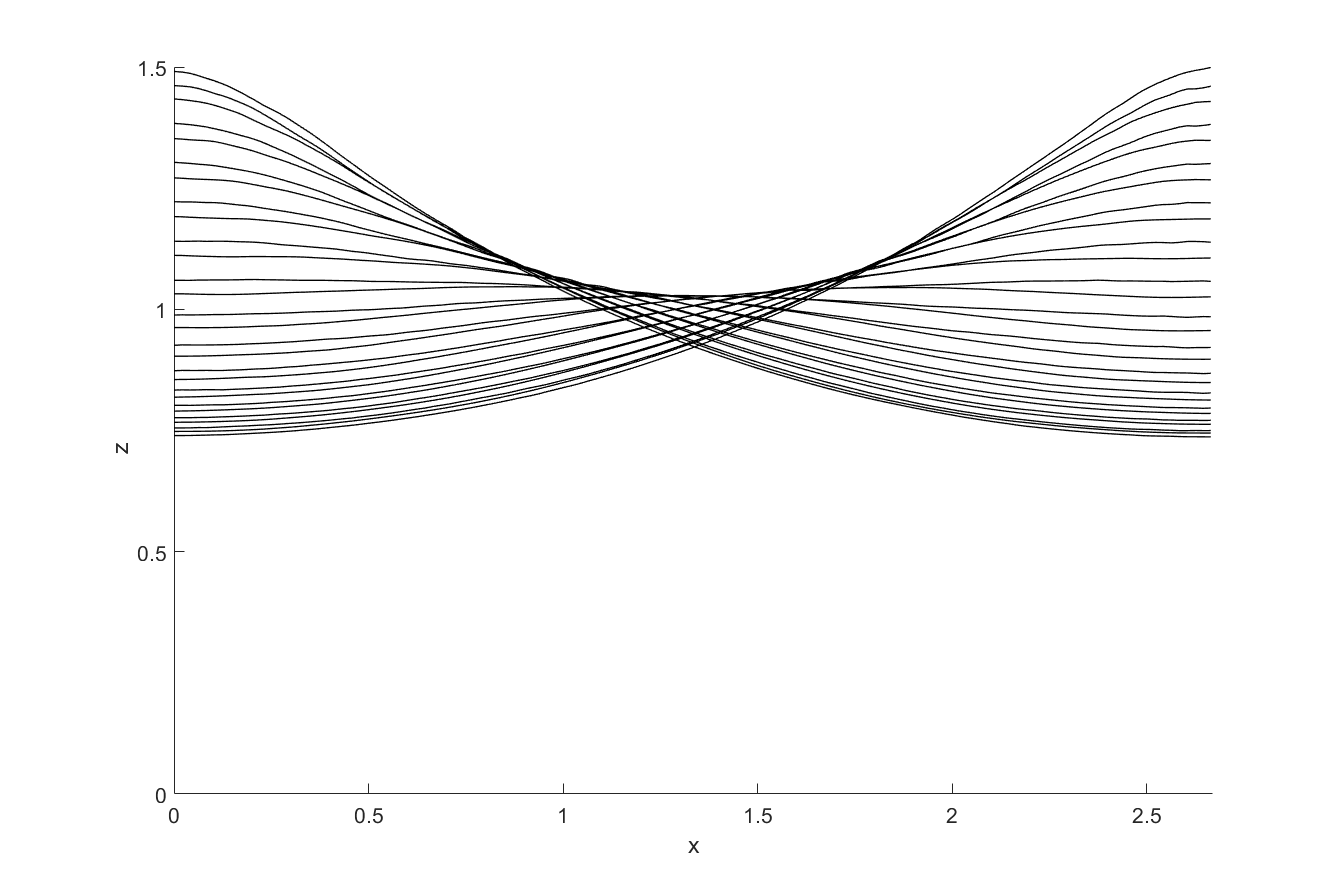}
        \includegraphics[width=0.45\textwidth]{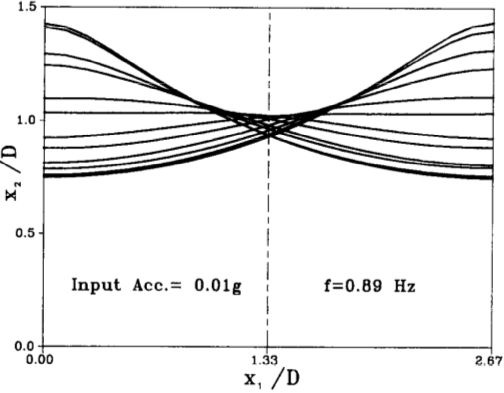}
\caption{Free surface mid-lines evolution : The computed results for 28 equally distributed time moments over one period (left) and for the purpose of comparison the reference results from \cite{huerta88} (right).}
\label{fig:surf_cmp}
\end{figure}

Figure \ref{fig:surf_cmp} demonstrates the same pattern  of the free surface evolution computed by the octree 3D code and the reference 2D results.

\subsection{Free surface flow passing rigidly mounted offshore oil platform}

To define the initial and boundary conditions for the simulation of sea waves passing a rigidly mounted obstacle, we consider simple, yet efficient, model of open sea waves  introduced in \cite{stokes_waves}  for the purpose of breaking waves animation.
The model is based on the third order Stokes wave which is defined as follows.

One starts by defining the first order Stokes wave in terms of $x$- and $z$-components of the free surface velocity $u,w$ and the water level $\eta$:
\begin{equation} \label{eq:first_order}
\begin{split}
\eta(x,t)& =  A\cos(kx-\omega t)\\
u(x,z,t)& =  A\omega e^{-kz}\cos(kx-\omega t)\\
w(x,z,t)& =  A\omega e^{-kz}\sin(kx-\omega t).
\end{split}
\end{equation}
Here $z=0$ is the mean water level,
$\omega = \frac{2\pi}T$ is the wave frequency,
$T$ is the wave period,
$k=\frac{2\pi}{\lambda}$ is the wave number,
$\lambda$ is the wave length.

Further one introduces the third order Stokes wave by the superposition of several first order Stokes waves  \eqref{eq:first_order}:
\begin{equation} \label{eq:third_order}
\begin{split}
\eta(x,t) =  \frac1{k}( \epsilon \cos(kx-\omega t) + \frac12\epsilon^2\cos(2kx-\omega t) + \frac38\epsilon^3\cos(3kx-\omega t)  )\\
u(x,z,t) =  \frac{\omega}{k}(  e^{-kz}\epsilon \cos(kx-\omega t) + \frac12 e^{-2kz}\epsilon^2\cos(2kx-\omega t) + \frac38 e^{-3kz}\epsilon^3\cos(3kx-\omega t)  )\\
w(x,z,t) =  \frac{\omega}{k}(  e^{-kz}\epsilon \sin(kx-\omega t) + \frac12 e^{-2kz}\epsilon^2\sin(2kx-\omega t) + \frac38 e^{-3kz}\epsilon^3\sin(3kx-\omega t)  ).\\
\end{split}
\end{equation}
With the help of \eqref{eq:third_order} we define the water level $\eta_{2D}(x,y,t)=\eta(x,t)$ and the bulk velocity
\[
\bu_{\rm wave}(x,y,z,t)=(u(x,z,t),0,w(x,z,t))^T\quad \text{for}~~z\le\eta_{2D}(x,y,t).
\]
We use $\bu_{\rm wave}(x,y,z,0)$ to prescribe the initial condition of our simulation.

The bulk computational domain is the  $440m\times 110m\times 110m$ box.  Box walls are orthogonal to the coordinate axes.
The sea depth is 55 m.
The inlet boundary is orthogonal to $x$-axis and has the minimal $x$-coordinate. The outlet  boundary  is opposite to the inlet  boundary.
On the inlet and outlet boundaries we impose the Dirichlet boundary condition using the Stokes wave,  $\bu_{\rm wave}(\bx,t)$, $\bx\in\Gamma_{in}\cup\Gamma_{out}$.
On other sides of the virtual box (except the top one) and the obstacle boundary we prescribe the no-penetration and free slip boundary condition.


\begin{figure}[!ht] \centering
\includegraphics[width=0.7\textwidth]{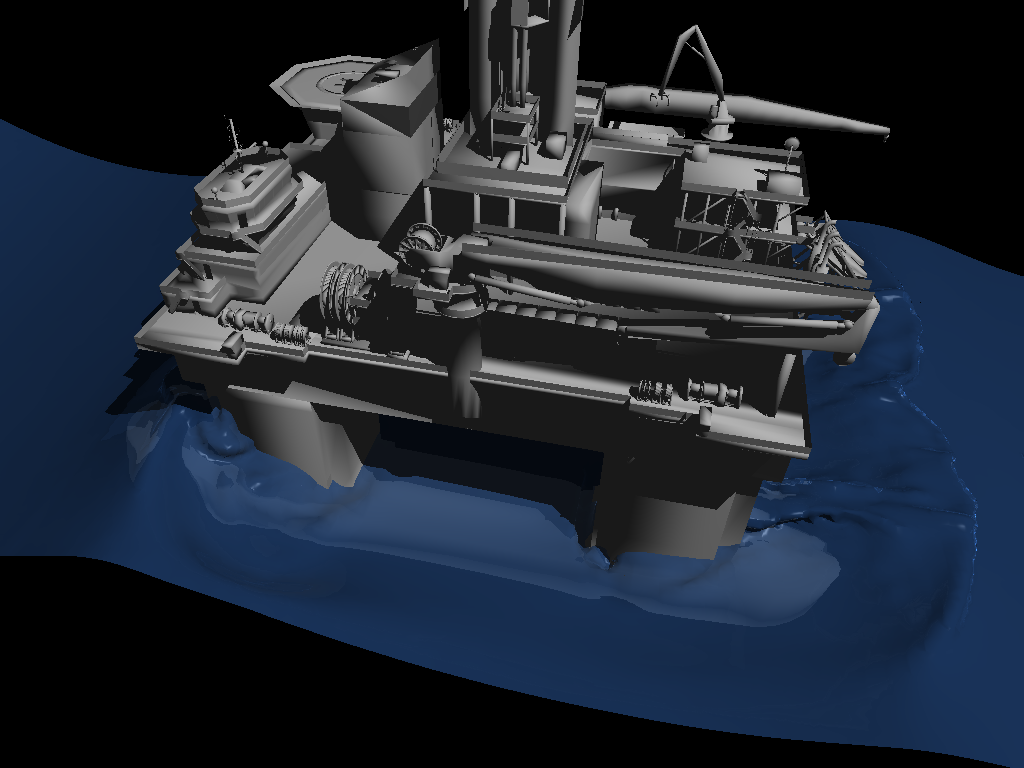}
\caption{A reconstruction of an operating offshore unit.}
\label{fig:pretty_platform}
\end{figure}

The partially submerged object of interest is a rigidly mounted offshore oil platform.
The platform shape is given by the reconstruction (with the help of a surface triangulation) of a currently operating  unit, see Figure \ref{fig:pretty_platform}.

The sea waves runup models the realistic weather scenario in the Kara sea offshore region.
In particular, $A=3$m and $T=4$s correspond to a moderate storm, whereas $A=11.5$m and $T=8.4$s define the largest waves recorded in this region over the time of observations.
In this paper we study the case of the largest sea waves with wave length $\lambda=110$m.
The practical statistics of interest are the highest water levels at the platform and forces experienced by the construction.

\begin{figure}[!ht] \centering
\includegraphics[width=0.8\textwidth]{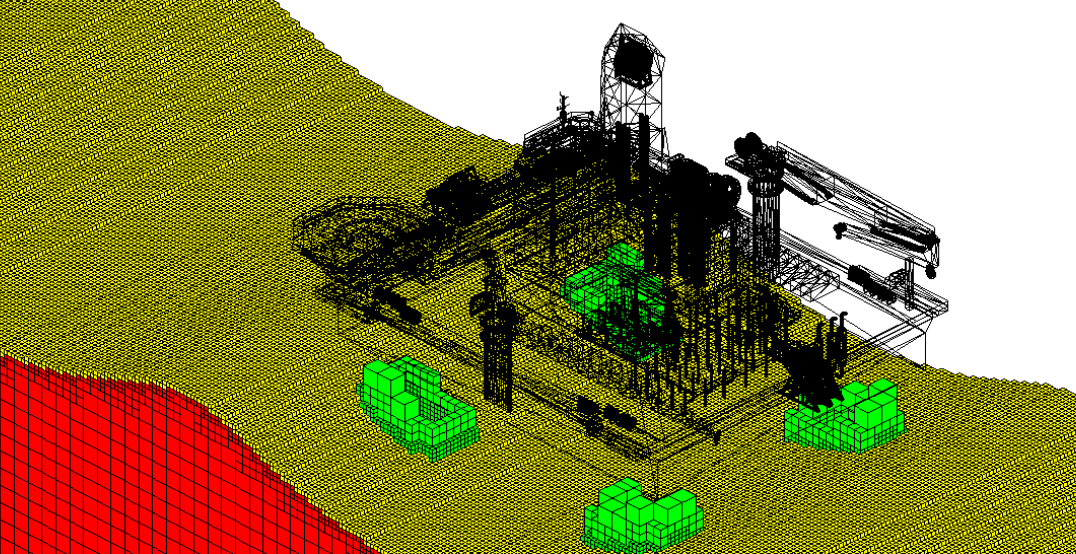}
\caption{Octree mesh for wave runup simulation: interior fluid cells (red), free surface cells (yellow), and immersed boundary cells (green).}
\label{fig:octree_platform}
\end{figure}

\begin{figure}[!ht] \centering
\includegraphics[width=0.8\textwidth]{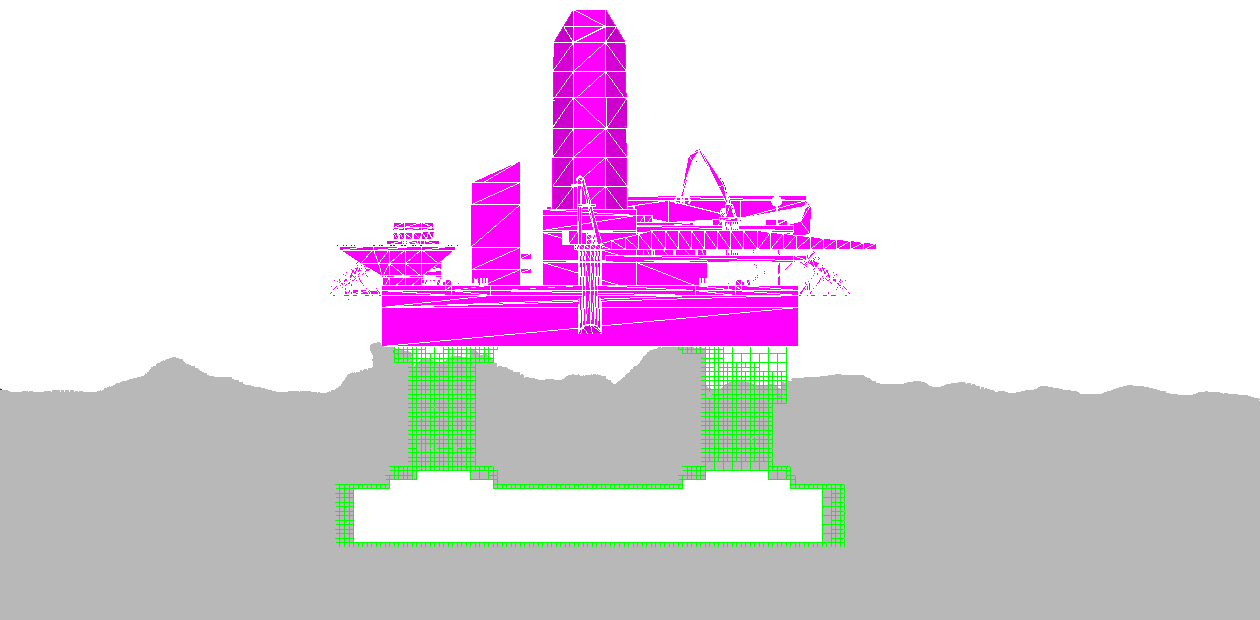}
\caption{Maximum observed water level, central cross-section of the computational domain.}
\label{fig:max_lev2}
\end{figure}

In Figure \ref{fig:octree_platform} we show the computational octree mesh, where different colors mark different type of cells: interior fluid, free surface, and solid boundary. {\color{black} We use the same dynamic  adaptation strategy as in the previous experiment with the sloshing tank.}
In Figure \ref{fig:max_lev2} we show the maximum water level observed in the simulation at the central cross-section of the computational domain.

\begin{figure}[!ht] \centering
\includegraphics[width=0.8\textwidth]{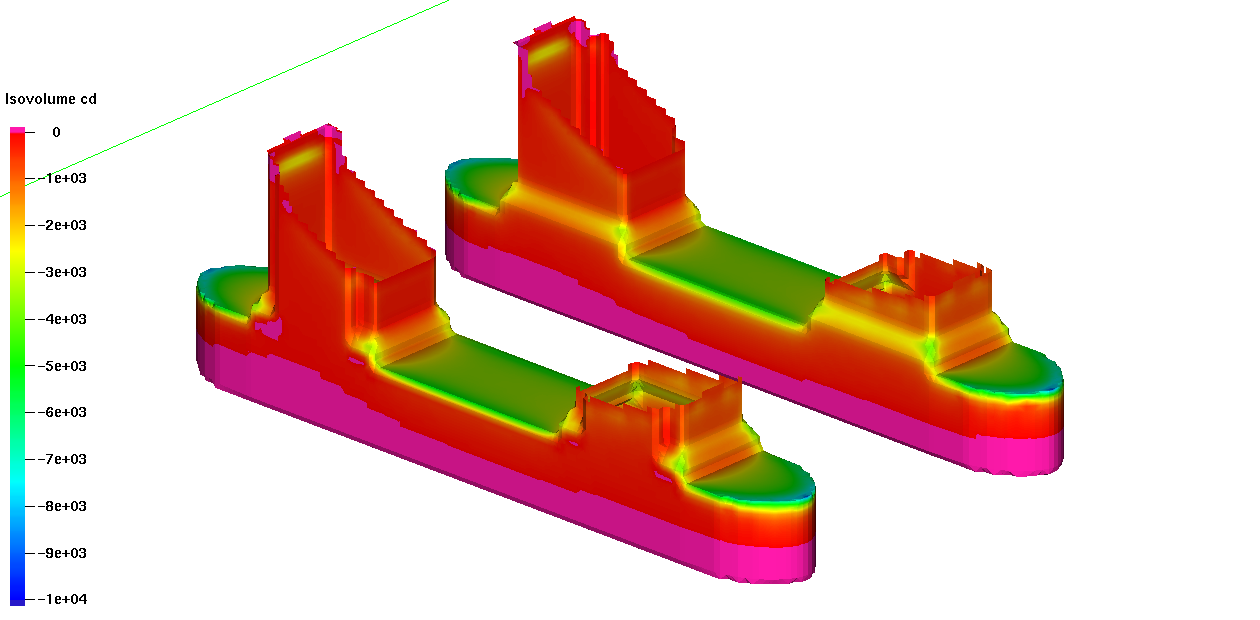}
\caption{A field of normal stresses projection at  $x$-direction}
\label{fig:subintegr}
\end{figure}

In Figure \ref{fig:subintegr} we present the maximum dynamic load experienced by the oilrig  piers in the x-direction.
The dynamic load is computed by taking the x-projection of the  normal stresses at the surface, i.e. the subintegral expression in \eqref{drag}.

\section{Conclusions}
We built a hybrid  finite volume\,/\,finite difference scheme for the simulation of  free-surface flows in complex geometries. The computational efficiency was achieved by using octree Cartesian meshes, while geometry  was handled through the immersing of both free and static boundaries in the background mesh. The major challenges were to construct compact stencil discretizations on the gradely refined meshes with low numerical dissipation and to enforce  various velocity and pressure boundary conditions on curvilinear parts of $\Gamma$. For a number of test examples, we demonstrated that the developed methods are particular suitable for the simulation of viscous free-surface flows over submerged or partially submerged objects.

\subsection*{Acknowledgements} The authors are grateful to N. Dianskiy and I. Kabatchenko for providing geophysical data for the Kara sea offshore.

\bibliographystyle{abbrv}
\bibliography{scheme}
\end{document}

%% file: inter_lin.pdf_t
\begin{picture}(0,0)%
\includegraphics[scale=0.48]{inter_lin.pdf}%
\end{picture}%
\setlength{\unitlength}{4144sp}%
\setlength{\unitlength}{.48\unitlength}%
\begingroup\makeatletter\ifx\SetFigFont\undefined%
\gdef\SetFigFont#1#2#3#4#5{%
  \reset@font\fontsize{#1}{#2pt}%
  \fontfamily{#3}\fontseries{#4}\fontshape{#5}%
  \selectfont}%
\fi\endgroup%
\begin{picture}(5444,5444)(-21,-4583)
\put(541,209){\makebox(0,0)[lb]{\smash{{\SetFigFont{12}{14.4}{\rmdefault}{\mddefault}{\updefault}{\color[rgb]{0,0,0}$\mathbf{x}_2$}%
}}}}
\put(361,-2041){\makebox(0,0)[lb]{\smash{{\SetFigFont{12}{14.4}{\rmdefault}{\mddefault}{\updefault}{\color[rgb]{0,0,0}$\mathbf{x}_1$}%
}}}}
\put(2251,-331){\makebox(0,0)[lb]{\smash{{\SetFigFont{12}{14.4}{\rmdefault}{\mddefault}{\updefault}{\color[rgb]{0,0,0}$\mathbf{x}_3$}%
}}}}
\put(2791,-4021){\makebox(0,0)[lb]{\smash{{\SetFigFont{12}{14.4}{\rmdefault}{\mddefault}{\updefault}{\color[rgb]{0,0,0}$\mathbf{x}_9$}%
}}}}
\put(991,-3526){\makebox(0,0)[lb]{\smash{{\SetFigFont{12}{14.4}{\rmdefault}{\mddefault}{\updefault}{\color[rgb]{0,0,0}$\mathbf{x}_{10}$}%
}}}}
\put(2071,-1636){\makebox(0,0)[lb]{\smash{{\SetFigFont{12}{14.4}{\rmdefault}{\mddefault}{\updefault}{\color[rgb]{0,0,0}$\mathbf{x}_V$}%
}}}}
\put(2971,-331){\makebox(0,0)[lb]{\smash{{\SetFigFont{12}{14.4}{\rmdefault}{\mddefault}{\updefault}{\color[rgb]{0,0,0}$\mathbf{x}_4$}%
}}}}
\put(4501,164){\makebox(0,0)[lb]{\smash{{\SetFigFont{12}{14.4}{\rmdefault}{\mddefault}{\updefault}{\color[rgb]{0,0,0}$\mathbf{x}_5$}%
}}}}
\put(1936,-2131){\makebox(0,0)[lb]{\smash{{\SetFigFont{12}{14.4}{\rmdefault}{\mddefault}{\updefault}{\color[rgb]{0,0,0}$\mathrm{y}$}%
}}}}
\put(4103,-2673){\makebox(0,0)[lb]{\smash{{\SetFigFont{12}{14.4}{\rmdefault}{\mddefault}{\updefault}{\color[rgb]{0,0,0}$\mathbf{x}_7$}%
}}}}
\put(4089,-3522){\makebox(0,0)[lb]{\smash{{\SetFigFont{12}{14.4}{\rmdefault}{\mddefault}{\updefault}{\color[rgb]{0,0,0}$\mathbf{x}_8$}%
}}}}
\put(4111,-1760){\makebox(0,0)[lb]{\smash{{\SetFigFont{12}{14.4}{\rmdefault}{\mddefault}{\updefault}{\color[rgb]{0,0,0}$\mathbf{x}_6$}%
}}}}
\end{picture}%

%% file: control_vol.pdf_t
\begin{picture}(0,0)%
\includegraphics[scale=0.56]{control_vol.pdf}%
\end{picture}%
\setlength{\unitlength}{4144sp}%
\setlength{\unitlength}{.56\unitlength}%
\begingroup\makeatletter\ifx\SetFigFont\undefined%
\gdef\SetFigFont#1#2#3#4#5{%
  \reset@font\fontsize{#1}{#2pt}%
  \fontfamily{#3}\fontseries{#4}\fontshape{#5}%
  \selectfont}%
\fi\endgroup%
\begin{picture}(5077,5216)(-554,-2555)
\put(-539,-601){\makebox(0,0)[lb]{\smash{{\SetFigFont{12}{14.4}{\rmdefault}{\mddefault}{\updefault}{\color[rgb]{0,0,0}$h$}%
}}}}
\put(2026,-2491){\makebox(0,0)[lb]{\smash{{\SetFigFont{12}{14.4}{\rmdefault}{\mddefault}{\updefault}{\color[rgb]{0,0,0}$\frac{h}{2}$}%
}}}}
\put(2971,-2491){\makebox(0,0)[lb]{\smash{{\SetFigFont{12}{14.4}{\rmdefault}{\mddefault}{\updefault}{\color[rgb]{0,0,0}$\frac{h}{4}$}%
}}}}
\end{picture}%

%% file: upw_points.pdf_t
\begin{picture}(0,0)%
\includegraphics[scale=0.5]{upw_points.pdf}%
\end{picture}%
\setlength{\unitlength}{4144sp}%
\setlength{\unitlength}{.5\unitlength}%
\begingroup\makeatletter\ifx\SetFigFont\undefined%
\gdef\SetFigFont#1#2#3#4#5{%
  \reset@font\fontsize{#1}{#2pt}%
  \fontfamily{#3}\fontseries{#4}\fontshape{#5}%
  \selectfont}%
\fi\endgroup%
\begin{picture}(6816,4273)(4018,-4751)
\put(8956,-2806){\makebox(0,0)[lb]{\smash{{\SetFigFont{12}{14.4}{\rmdefault}{\mddefault}{\updefault}{\color[rgb]{0,0,0}$r$}%
}}}}
\put(7201,-1861){\makebox(0,0)[lb]{\smash{{\SetFigFont{12}{14.4}{\rmdefault}{\mddefault}{\updefault}{\color[rgb]{0,0,0}Advective flux nodes}%
}}}}
\put(7201,-1411){\makebox(0,0)[lb]{\smash{{\SetFigFont{12}{14.4}{\rmdefault}{\mddefault}{\updefault}{\color[rgb]{0,0,0}Reference nodes}%
}}}}
\put(7201,-961){\makebox(0,0)[lb]{\smash{{\SetFigFont{12}{14.4}{\rmdefault}{\mddefault}{\updefault}{\color[rgb]{0,0,0}Flow direction}%
}}}}
\put(9811,-2806){\makebox(0,0)[lb]{\smash{{\SetFigFont{12}{14.4}{\rmdefault}{\mddefault}{\updefault}{\color[rgb]{0,0,0}$H$}%
}}}}
\put(9586,-3841){\makebox(0,0)[lb]{\smash{{\SetFigFont{12}{14.4}{\rmdefault}{\mddefault}{\updefault}{\color[rgb]{0,0,0}$h$}%
}}}}
\put(8371,-4696){\makebox(0,0)[lb]{\smash{{\SetFigFont{12}{14.4}{\rmdefault}{\mddefault}{\updefault}{\color[rgb]{0,0,0}$\Delta x$}%
}}}}
\put(5896,-3571){\makebox(0,0)[lb]{\smash{{\SetFigFont{12}{14.4}{\rmdefault}{\mddefault}{\updefault}{\color[rgb]{0,0,0}$u_{-1}$}%
}}}}
\put(8596,-3571){\makebox(0,0)[lb]{\smash{{\SetFigFont{12}{14.4}{\rmdefault}{\mddefault}{\updefault}{\color[rgb]{0,0,0}$u_0$}%
}}}}
\put(9946,-3571){\makebox(0,0)[lb]{\smash{{\SetFigFont{12}{14.4}{\rmdefault}{\mddefault}{\updefault}{\color[rgb]{0,0,0}$u_1$}%
}}}}
\put(10576,-3571){\makebox(0,0)[lb]{\smash{{\SetFigFont{12}{14.4}{\rmdefault}{\mddefault}{\updefault}{\color[rgb]{0,0,0}$u_2$}%
}}}}
\end{picture}%

%% file: ref_points_dif.pdf_t
\begin{picture}(0,0)%
\includegraphics[scale=0.55]{ref_points_dif.pdf}%
\end{picture}%
\setlength{\unitlength}{4144sp}%
\setlength{\unitlength}{.55\unitlength}%
\begingroup\makeatletter\ifx\SetFigFont\undefined%
\gdef\SetFigFont#1#2#3#4#5{%
  \reset@font\fontsize{#1}{#2pt}%
  \fontfamily{#3}\fontseries{#4}\fontshape{#5}%
  \selectfont}%
\fi\endgroup%
\begin{picture}(6816,3666)(4018,-4144)
\put(7201,-1861){\makebox(0,0)[lb]{\smash{{\SetFigFont{12}{14.4}{\rmdefault}{\mddefault}{\updefault}{\color[rgb]{0,0,0}Diffusion flux nodes}%
}}}}
\put(7201,-1411){\makebox(0,0)[lb]{\smash{{\SetFigFont{12}{14.4}{\rmdefault}{\mddefault}{\updefault}{\color[rgb]{0,0,0}Reference nodes}%
}}}}
\put(9811,-2806){\makebox(0,0)[lb]{\smash{{\SetFigFont{12}{14.4}{\rmdefault}{\mddefault}{\updefault}{\color[rgb]{0,0,0}$H$}%
}}}}
\put(9586,-3841){\makebox(0,0)[lb]{\smash{{\SetFigFont{12}{14.4}{\rmdefault}{\mddefault}{\updefault}{\color[rgb]{0,0,0}$h$}%
}}}}
\put(5896,-3571){\makebox(0,0)[lb]{\smash{{\SetFigFont{12}{14.4}{\rmdefault}{\mddefault}{\updefault}{\color[rgb]{0,0,0}$\mathbf{x}_{-1}$}%
}}}}
\put(9946,-3571){\makebox(0,0)[lb]{\smash{{\SetFigFont{12}{14.4}{\rmdefault}{\mddefault}{\updefault}{\color[rgb]{0,0,0}$\mathbf{x}_1$}%
}}}}
\put(10576,-3571){\makebox(0,0)[lb]{\smash{{\SetFigFont{12}{14.4}{\rmdefault}{\mddefault}{\updefault}{\color[rgb]{0,0,0}$\mathbf{x}_2$}%
}}}}
\put(8911,-3841){\makebox(0,0)[lb]{\smash{{\SetFigFont{12}{14.4}{\rmdefault}{\mddefault}{\updefault}{\color[rgb]{0,0,0}$r$}%
}}}}
\put(7201,-2806){\makebox(0,0)[lb]{\smash{{\SetFigFont{12}{14.4}{\rmdefault}{\mddefault}{\updefault}{\color[rgb]{0,0,0}$R$}%
}}}}
\put(8101,-3571){\makebox(0,0)[lb]{\smash{{\SetFigFont{12}{14.4}{\rmdefault}{\mddefault}{\updefault}{\color[rgb]{0,0,0}$\mathbf{x}_0$}%
}}}}
\end{picture}%

%% file: oct-curv-1.pdf_t
\begin{picture}(0,0)%
\includegraphics[scale=0.5]{oct-curv-1.pdf}%
\end{picture}%
\setlength{\unitlength}{4144sp}%
\setlength{\unitlength}{.5\unitlength}%
\begingroup\makeatletter\ifx\SetFigFont\undefined%
\gdef\SetFigFont#1#2#3#4#5{%
  \reset@font\fontsize{#1}{#2pt}%
  \fontfamily{#3}\fontseries{#4}\fontshape{#5}%
  \selectfont}%
\fi\endgroup%
\begin{picture}(10910,6382)(1253,-4621)
\put(5266,-3256){\makebox(0,0)[lb]{\smash{{\SetFigFont{12}{14.4}{\rmdefault}{\mddefault}{\updefault}{\color[rgb]{0,0,0}$\mathbf{x}_2$}%
}}}}
\put(3286,-1816){\makebox(0,0)[lb]{\smash{{\SetFigFont{12}{14.4}{\rmdefault}{\mddefault}{\updefault}{\color[rgb]{0,0,0}$\mathbf{x}_1$}%
}}}}
\put(9901,-3661){\makebox(0,0)[lb]{\smash{{\SetFigFont{12}{14.4}{\rmdefault}{\mddefault}{\updefault}{\color[rgb]{0,0,0}Internal d.o.f.}%
}}}}
\put(9901,-2761){\makebox(0,0)[lb]{\smash{{\SetFigFont{12}{14.4}{\rmdefault}{\mddefault}{\updefault}{\color[rgb]{0,0,0}Boundary d.o.f.}%
}}}}
\put(9901,-1861){\makebox(0,0)[lb]{\smash{{\SetFigFont{12}{14.4}{\rmdefault}{\mddefault}{\updefault}{\color[rgb]{0,0,0}Inactive d.o.f.}%
}}}}
\put(9901,-961){\makebox(0,0)[lb]{\smash{{\SetFigFont{12}{14.4}{\rmdefault}{\mddefault}{\updefault}{\color[rgb]{0,0,0}Internal cell center}%
}}}}
\put(9901,-61){\makebox(0,0)[lb]{\smash{{\SetFigFont{12}{14.4}{\rmdefault}{\mddefault}{\updefault}{\color[rgb]{0,0,0}Boundary cell center}%
}}}}
\end{picture}%